\documentclass[prb, 10pt, twocolumn, aps, showpacs,preprintnumbers,amsmath,amssymb, longbibliography]{revtex4-1}

\usepackage{graphicx}
\usepackage{dcolumn}
\usepackage{bm}
\usepackage[bookmarks, colorlinks=true, breaklinks]{hyperref}
\hypersetup{linkcolor=blue,citecolor=blue,filecolor=black,urlcolor=blue}
\usepackage{color}

\emergencystretch=\maxdimen
\newcommand{\bra}{\left< }
\newcommand{\ket}{\right>}

\begin{document}

\title{Quantum ferroelectric instabilities in superconducting SrTiO$_3$} 

\author{J. R. Arce-Gamboa$^1$,  G. G. Guzm\'{a}n-Verri$^{1,2}$}
\email[Corresponding author: ]{gian.guzman@ucr.ac.cr}
\affiliation{$^{1}$Centro de Investigaci\'{o}n en Ciencia e Ingenier\'{i}a de Materiales (CICIMA), Universidad de Costa Rica, San Jos\'{e}, Costa Rica 11501}
\affiliation{$^{2}$Materials Science Division, Argonne National Laboratory, Argonne, Illinois, USA 60439,}

\date{\today}

\begin{abstract}
We examine the effects of strain and cation substitution on the superconducting phase of  polar semiconductors near a  ferroelectric quantum phase transition with a
model that combines a strong coupling theory of superconductors
with a standard microscopic framework for displacive polar modes coupled to strain degrees of freedom. 
Our calculations reveal that the superconducting transition temperature $T_c$ is enhanced by proximity to the ferroelectric instability from the disordered side, while it is 
generally suppressed in the ordered phase due to its increase in dielectric stiffness and a reduction 
of critical fluctuations from dipolar induced anisotropies. 
The condensation of the  pairing phonon excitations generates a kink in $T_c$ 
at a charge density that is generally lower than that of the quantum critical point (QCP) and where both superconducting and ferroelectric orders set in.
We apply our model to SrTiO$_3$ and find that the antiadiabatic limit 
places the kink nearly at its QCP. 
As the QCP is pushed to higher charge densities with either tuning parameter,
we find that the dome narrows and sharpens. 
Our model is in qualitative and fair quantitative agreement
with the recent observation of 
overlapping ferroelectric-like
and superconducting instabilities 
in n-doped Sr$_{1-x}$Ca$_x$TiO$_3$
and strain tuning of $T_c$ in n-doped SrTiO$_3$.
We compare our results to  previous models invoking order-disorder lattice dynamics to
describe the pairing excitations.
\end{abstract}

\maketitle

\section{Introduction}

Strontium titanate (STO)  is the first superconducting~(SC) oxide to be discovered more 
than half a century ago~\cite{Schooley1964a}. Upon doping, 
it exhibits dome-shaped superconductivity~\cite{Schooley1965a, Koonce1967a, Binning1980a, Suzuki1996a} similar to the well-known cuprates
at unusually low charge carrier concentrations ($< 10^{18}\,$cm$^{-3}$)~\cite{Eagles1986a, Eagles1989a, Tainsh1986a, Eagles2016a},
 dubbing STO as “the most dilute superconductor”~\cite{Lin2013a}.
Since its observation, the origin of the pairing mechanism between its charge carriers has posed a long-standing problem in condensed matter physics, 
as the conventional Bardeen-Cooper-Schrieffer theory of superconductors
 requires  the so-called adiabatic limit of electron-phonon coupling in which the
characteristic energy scale of the charge carriers (the Fermi temperature $T_F$) is much larger than
that of the phonons (the Debye temperature $T_D$)~\cite{Cabrotte1990a}.
STO is unusual in that 
it is outside this limit as
$T_F \simeq 13\,$K~\cite{Lin2013a} 
and $T_D \simeq 400\,$K~\cite{Burns1980a}.
Though several models have been put forward~\cite{Gorkov2016a, Ruhman2016a, Takada1980a, Edelman2017a},
and experiments have constrained possible pairing mechanisms between its charge carriers~\cite{Swartz2018a},
there is still no consensus on a theory of superconductivity in dilute polar semiconductors such as doped STO~\cite{Collignon2018a}.

Somewhat recently, Rowley \textit{et al}.~\cite{Rowley2014a} have suggested a connection between quantum criticality and 
superconductivity in STO, and 
 Edge \textit{et al}.~\cite{Edge2015a} have put forward a concrete model. Within this framework, 
a QCP
 that separates competing ferroelectric (FE)
 and paraelectric (PE) ground states 
generates highly collective low energy phonon excitations
which induce an instability in the metallic phase and
pair the charge carriers to form Cooper pairs.
 A dome arises because the quantum FE fluctuations pair up the charge carriers 
at low doping levels, though there are not enough of them to 
provide a robust superconductivity. As the density increases, so does $T_c$ until the fluctuations
become ineffective at pairing due to screening by the carriers themselves. 
Though STO is not within the adiabatic limit,
the model  provides a good description of its SC dome,
and it has received recent experimental support~\cite{Rischau2017a, Stucky2016a, Herrera2018a}.

In conventional FEs, the relevant low energy lattice excitations are zone-center transverse optic (TO) phonons
which break spatial inversion symmetry and trigger a phase with a spontaneous, reversible polarization 
upon condensation. In pure STO, 
such a transition is aborted by quantum 
fluctuations of the polarization order parameter~\cite{Muller1979a} and,  
instead, incipient FE behavior 
is observed  in which the dielectric constant grows enormously ($\simeq 10^{4}$) and the phonon frequencies
soften down to lowest observed temperatures without condensing~\cite{Yamada1969a}.
Long-range FE order has been induced by tensile stress~\cite{Hiromoto1976a}, oxygen 
isotope exchange~\cite{Itoh1999a}, or cation substitution~\cite{Bednorz1984a}. 
 At the QCP, the phonon excitations become gapless, and 
at finite temperatures above it but well below $T_D$~\cite{Chandra2017a},  quantum criticality sets in
generating unusual behavior,  
such as a $T^{-2}$  dependence in the dielectric susceptibility 
distinct from that of the neighboring PE and FE phases~\cite{Rowley2014a}.

To model the pairing excitations between carriers, 
one-dimensional transverse Ising models with short-range interactions
have been invoked and solved within a mean field approximation~\cite{Edge2015a, Kedem2016a, Dunnett2016a}.
However, the 
soft collective vibrational excitations described by such models are not normal 
modes of the lattice but rather unstable pseudospin waves with relaxational dynamics,
which are typical of order-disorder~(OD) FEs such as potassium dihydrogen phosphate (KH$_2$PO$_4$)~\cite{Lines2001a}.
STO, on the other hand, is a perovskite with a low temperature tetragonal PE phase which
exhibits displacive behavior with a resonant soft 
TO mode that fully accounts for the dielectric constant and remains undamped down to the lowest observed temperatures~\cite{Rowley2014a, Yamada1969a, Inoue1983a},
even when the lattice instability is approached by oxygen 
isotope substitution where it has been suggested that the effects of an OD component 
may become important~\cite{Blinc2005a, Takesada2006a, *Taniguchi2007a, *Yagi2009a}.
Moreover,  FE transitions in perovskites are generally understood as resulting from the competition between 
a PE state favored by short-range repulsive forces  and a FE phase favored by long-ranged and anisotropic dipole interactions
that arise from their mixed covalent-ionic bonding character.~\cite{ModernPerspective} 
The aim of this paper is to describe the SC phase that emerges from 
a pairing excitation that corresponds to a displacive soft TO phonon.

We use a self-consistent phonon approximation~(SCPA) 
to study the quantum statistical mechanics of our model Hamiltonian which
includes local thermal and quantum fluctuations of the order parameter. This
allows us to self-consistently calculate the doping and strain dependence of $T_c$, as well 
the FE transition temperature $T_{fe}$, phonon energies and polarization order parameter. 
We neglect the effects of band narrowing due to polaron formation, 
which could have strong effects on the correlations~\cite{Gabay2017a, *Wang2016a, *Marel2011a, *Devreese2010a, *Meevasana2010a, *Mechelen2008a}.

We find that proximity to the QCP  favors superconductivity from the PE side, but  
$T_c$ decreases on the ordered side due to a reduction of the critical fluctuations from dipolar induced anisotropies and 
the dielectric stiffening by the spontaneous polarization. 
This generates a kink in the SC dome at a characteristic charge density that is generally lower than that of the QCP
and where the SC and FE orders set in.
As the QCP is pushed to higher charge densities with cation substitution or tensile strain
so the FE phase increasingly overlaps with the SC phase, the dome narrows and sharpens. 
Though we do not make any \textit{a-priori} assumptions that the pairing fluctuations are quantum critical,
we find that the antiadiabatic limit places the kink nearly at the QCP and within the QC region.
While our results are in qualitative agreement with the essentials of the OD models,
there are significant qualitative differences. 
Though our model produces a dome which is narrower than the observed one,
its phase diagram is in overall 
 agreement with recent experiments 
in n-doped Sr$_{1-x}$Ca$_x$TiO$_3$ (STO:Ca-$x$)~\cite{Rischau2017a}
and the observed strain tuning of $T_c$
in n-doped STO~\cite{Rowley2018a, Pfeiffer1970a, Herrera2018a} 

\section{Model}

Following Ref.~[\onlinecite{Edge2015a}],
our starting point is the McMillian formula~\cite{McMillian1968a} for the SC 
coupling constant,
\begin{align*}
	\lambda(n,T, P) = \int_0^\infty \alpha_{e-ph}^2 (\omega) F(\omega) \frac{d \omega}{\omega},
\end{align*}
where $\alpha_{e-ph}(\omega)$ is the electron-phonon coupling,   
 $F(\omega)= \sum_{{\bm q},\mu} \delta \left( \omega - \Omega_{{\bm q} \mu} \right) $,
is the spectral density at frequency $\omega$, and 
$\Omega_{{\bm q} \mu}\, (\mu=1,2, 3)$ is the phonon energy at 
wave vector  ${\bm q}$.
Generally, it is not expected that charge carriers couple to long-wavelength TO phonons.~\cite{Giustino2017a}
For perovskite lattices such as that of STO, this is indeed the case but only when ${\bm q}$ is along a principal axis.
Away from such special directions, coupling occurs due to cubic anisotropy.~\cite{Woelfe2018a}
While such finite coupling still depends on $\Omega_{{\bm q} \mu}$,
we make the  simplifying assumption that 
$\alpha_{e-ph}$ is independent of it. Thus,
\begin{align}
	\label{eq:lambda}
	\lambda(n,T, P) = \alpha_{e-ph}^2 \sum_{{\bm q},\mu} \frac{1}{ \Omega_{{\bm q} \mu} },
\end{align}
where the sum over ${\bm q}$ runs over the entire Brillouin zone.  
We note that despite this assumption
 $\lambda$ does not diverge:
The largest contributions occur at the FE transition where the TO phonons become gapless 
and have a dispersion $\Omega_{\bm q} \propto q$ [see Eq.(\ref{eq:TOLOCb}) below]. By taking the 
continuum limit of Eq.~(\ref{eq:lambda})  over a sphere of wave vector cutoff radius $\Lambda$,
we find that
$\lambda \propto \int d^3{\bm q} / \Omega_{\bm q} \propto  \int_0^{\Lambda} dq q^{2} / q \propto \Lambda^2$.

$T_{c}$ is calculated from the strong-coupling theory,~\cite{Edge2015a}
\begin{align}
	\label{eq:Tsc}
	1 = \frac{\lambda(n, T_c,P)}{2 \pi^2} \int_{-\epsilon_F}^0 d \epsilon N(\epsilon) \frac{\tanh( \beta_{c} \epsilon /2  ) }{\epsilon},
\end{align}
where $N(\epsilon) \simeq	 \sqrt{\epsilon + \epsilon_F}$ is the electron density of states near the Fermi level $\epsilon_F$,
and $\beta_{c} = (k_B T_{c})^{-1}$.

We now need a model for  the phonon excitations. 
We consider a standard 
model Hamiltonian for displacive FEs
with  normal mode coordinates that describe local displacements 
$ {\bm Q}_i  = ( Q_{ix}, Q_{iy}, Q_{iz})$   in the unit cell $i$  that are 
associated with the soft TO mode, the condensation of which is driven by 
the dipolar force and leads to the FE transition~\cite{Lines2001a}. 
We also consider elastic strains $\eta_{\alpha}$  coupled
to the displacements ${\bm Q}_i$.
We  write the components of the strain tensor  in the usual Voigt notation:
$ \eta_1  \equiv  \epsilon_{xx}$, $ \eta_2  \equiv  \epsilon_{yy}$, $ \eta_3  \equiv  \epsilon_{zz}$,  $ \eta_4 = 2\epsilon_{yz} ,  \eta_5 = 2\epsilon_{xz} $, and  $\eta_6 = 2\epsilon_{xy}$.

The  Hamiltonian is as follows~\cite{Pytte1972a},
\begin{align}
	\label{eq:Hamiltonian}
	H = H_Q + H_\eta + H_{Q \eta},
\end{align}
where,
\begin{subequations}
\begin{align}
\label{eq:HamiltonianQ}
H_Q &=    \frac{1}{2} \sum_{i  }   \left| {\bm \Pi}_{ i } \right|^2 + \frac{\kappa}{2} \sum_{ i }  \left| {\bm Q}_{ i } \right|^2
+ \frac{\alpha}{4} \sum_{ i } \left| {\bm Q}_{ i } \right| ^4 \nonumber \\
&~~~~~ + \frac{ \gamma}{ 2 } \sum_{i, \nu \neq \nu^\prime }  Q_{ i \nu }^2  Q_{ i \nu^\prime }^2 
-\frac{1}{2} \sum_{ij \nu \nu^\prime} v_{ij}^{\nu \nu^\prime}  Q_{ i \nu }   Q_{ j \nu^\prime },  
\end{align}
\begin{align}
	H_\eta &=  \frac{1}{2} \sum_{i, \mu \mu^\prime=1}^6 C_{\mu \mu^\prime} \eta_{\mu i} \eta_{\mu^\prime i} + P \sum_{i, \mu=1}^3 \eta_{\mu i},
\end{align}
\begin{align}
	\label{eq:HamiltonianQeta}
	H_{Q\eta}  &=  -  e_a  \sum_{ i } ( \eta_{1i} + \eta_{2i} + \eta_{3i})  \left| {\bm Q}_{ i } \right|^2 \\  \nonumber 
       &- e_t \sum_i \left[ \eta_{1i} \left( 2 Q_{ix}^2-Q_{iy}^2-Q_{iz}^2 \right) \right. \\ \nonumber 
       &~~~~~~~~~~~~~~~~
       + \eta_{2i} \left( 2 Q_{iy}^2-Q_{ix}^2-Q_{iz}^2  \right)  \\ \nonumber 
       &~~~~~~~~~~~~~~~~~~~~~~~~~~~~~~~
       \left. + \eta_{3i} \left( 2 Q_{iz}^2-Q_{ix}^2-Q_{iy}^2  \right) \right] \\ \nonumber
       & - e_r \sum_i ( Q_{i x} Q_{i y} \eta_{6i} + Q_{i x} Q_{i z} \eta_{5i} +Q_{i y} Q_{i z} \eta_{4i}).
\end{align}
\end{subequations}
Here, $ {\bm \Pi}_i $ is  
the conjugate momentum of $  {\bm Q}_i  $, and 
$ v_{ij}^{\nu \nu^\prime} $~($\nu,\nu^\prime = x,y,z$) is the dipolar interaction tensor with 
Fourier transform
$
v_{\bm q}^{ \nu \nu^\prime } 
=  \left[ \frac{1}{3}C^2 - B^2 q^2  \right] \delta_{ \nu \nu^\prime } -C^2 \frac{ q_{ \nu } q_{ \nu^\prime } }{ q^2 },  
$
where $ q = \left| {\bm q} \right| $ and  $B$ and $C$ 
are constants that depend on the lattice structure~\cite{Aharony1973a}.
$\kappa$ is the lattice stiffness; $\alpha$ and $ \gamma$ are 
coefficients of
the isotropic and anisotropic cubic anharmonicities, respectively.
$e_a, e_t,$ and $e_r$ are coupling constants between the polar and strain degrees of freedom; 
$C_{\mu \mu^\prime}$~($\mu,\mu^\prime = 1,2,...,6$) is the elastic constant tensor in the cubic phase, 
and $P$ is a hydrostatic pressure, both in units of energy~ (the
usual elastic constants  and homogeneous stresses are 
given by $C_{\mu \mu^\prime} a^{-3}$ and $P a^{-3}$ where $a \simeq 3.9\,$\AA\,  is the lattice constant of the cubic structure).
We consider fictitious negative hydrostatic pressures 
as a simple way to model the effects of tensile strain on the superconducting dome. 
Tensile strains are of course the experimentally relevant cases~\cite{Hiromoto1976a}.
We study the quantum statistical mechanics of the Hamiltonian~(\ref{eq:Hamiltonian}) 
within SCPA~\cite{Pytte1972a}. We consider the PE and FE phases separately.

\subsection{PE phase}
For simplicity, we will assume that the PE phase is cubic.
Above $T_{fe}$, there is therefore a
 doubly degenerate TO phonon $ \Omega_{ {\bm q}}^\perp $ and a singlet longitudinal optic (LO) mode 
$  \Omega_{ {\bm q}}^\parallel $ with isotropic dispersions~\cite{Arce-Gamboa2017a},
\begin{subequations}
\label{eq:TOLOC}
\begin{align}
\label{eq:TOLOCa}
\left( \Omega_{ {\bm q}}^{\parallel} \right)^2  &= \left( \Omega_{\bm q}^{\perp} \right)^2  + C^2, \\
\label{eq:TOLOCb}
 \left( \Omega_{ {\bm q}}^{\perp} \right)^2 &=  \left( \Omega_{0}^{\perp} \right)^2 + B^2 q^2,
\end{align}
\end{subequations}
where $\Omega_{0}^{\perp}$ is the TO mode at the zone center.  
As expected~\cite{LevanyukFerroelectricity}, 
the effect of the dipole force is to lift the triply degenerate mode of the cubic phase by 
 gapping out the LO phonons.

Within SCPA,  $\Omega_{0}^{\perp}$ is given as follows~\cite{Pytte1972a},
\begin{align}
\label{eq:TOLOC0}
 \left( \Omega_{0}^{\perp} \right)^2  &=  \omega_0^2 + \left( 5 \alpha + 2 \gamma  \right)  \psi_{0} - 2e_a \eta_a,
\end{align}
where $\omega_0 \equiv \sqrt{ \kappa - v_0 }$, is the frequency of a purely harmonic
model and $v_0 \equiv C^2/3$ is the largest Fourier component of the dipole interaction; 
$\psi_0 =  \left( 2  \omega \right)^{-1} \coth\left( \beta \omega / 2 \right)$ are local fluctuations of polarization 
with $  \omega = \sqrt{ \left( \Omega_{0}^{\perp} \right)^2 + v_0} $ and $\eta_a = \bra \eta_1 + \eta_2 + \eta_3 \ket =  (e_a/C_a) \left( 3 \psi_0 \right) - P /C_a$ is the volume strain. $\bra ... \ket$ denotes thermal average. 

According to Eqs.~(\ref{eq:lambda}) and (\ref{eq:TOLOC}), we therefore have,
\begin{align}
	\label{eq:lambdaPE}
	\lambda(n, T_{c},P) = \alpha_{e-ph}^2   \sum_{\bm q} \left( \frac{2}{ \Omega_{ {\bm q}}^{\perp} } +   \frac{1}{ \Omega_{ {\bm q}}^{\parallel} } \right).
\end{align}
Note that the largest contributions to $\lambda$ come from the critical mode $\Omega_{ {\bm q}}^{\perp}$,
as  $\Omega_{ {\bm q}}^{\parallel}$ is gapped out by the large depolarizing field. 

We now parametrize the model parameters 
in order to describe the effects of doping and cation substitution on the phonon excitations.
In the OD models~\cite{Edge2015a, Kedem2016a, Dunnett2016a}, 
the quantum tunneling energy  $\Gamma$  between FE ground states with opposite polarization
is  chosen for such parametrization, as the effects of quantum fluctuations become important
when $\Gamma$ is comparable to the well-depth between such states. 
For displacive models, however, 
quantum fluctuations must be comparable to the structural
differences between competing PE and FE ground states and
generate zero-point energies 
comparable to the classical energy reduction~\cite{Zhong1996a}. 
We therefore choose to parametrize such energy barrier, which in our model is proportional to $-\omega_0^4/ \alpha$  at $0\,$K. 
For simplicity, we keep $\alpha$ fixed and assume that all changes occur in $\omega_0$  as follows,
\begin{align}
	\label{eq:w0}
	\omega_0^2 \to 	\omega_0^2\left[ 1  - b_2 \left( e^{ \epsilon_F / b_1} - 1 \right) -   g (x_r - x)  \right],
\end{align}
where $b_1, b_2, g$ are model parameters that will be fitted to experiments. 
$x_r = 0.018$ is the Ca concentration above which 
STO:Ca-$x$ enters a glassy phase  
which we do not aim to describe here~\cite{Bednorz1984a}.
This parametrization is constructed based on (i) the observation that doping destabilizes
the FE phase~\cite{Rischau2017a, Edge2015a} while cation substitution stabilizes  it;
and (ii) the restriction that our model should generate simultaneous physically 
reasonable values for  $T_c$ and $T_{fe}$ as well as $\Omega_{\bm q}$ in the relevant doping range. 
We will show below that Eq.~(\ref{eq:w0}) accomplishes this at the expense of narrowing the SC dome compared to the observed one. 
We have attempted to use the polynomial parametrization of the OD models,
and while we can find a set of model parameters that fit the observed SC dome,
 we could not obtain physical values for $T_{fe}$  and $\Omega_{\bm q}$. 
 Clearly, this highlights the need for a theory to model these effects~\cite{Woelfe2018a}.

Equations~(\ref{eq:Tsc}) and (\ref{eq:TOLOC})-(\ref{eq:w0})  
are a self-consistent system that give 
 $T_{c}(n, P)$, $T_{fe}(n, P)$, and $\Omega_{0}^{\perp}(n, P, T)$. 
\begin{figure}[htp]
	\centering
	\includegraphics[scale=0.5]{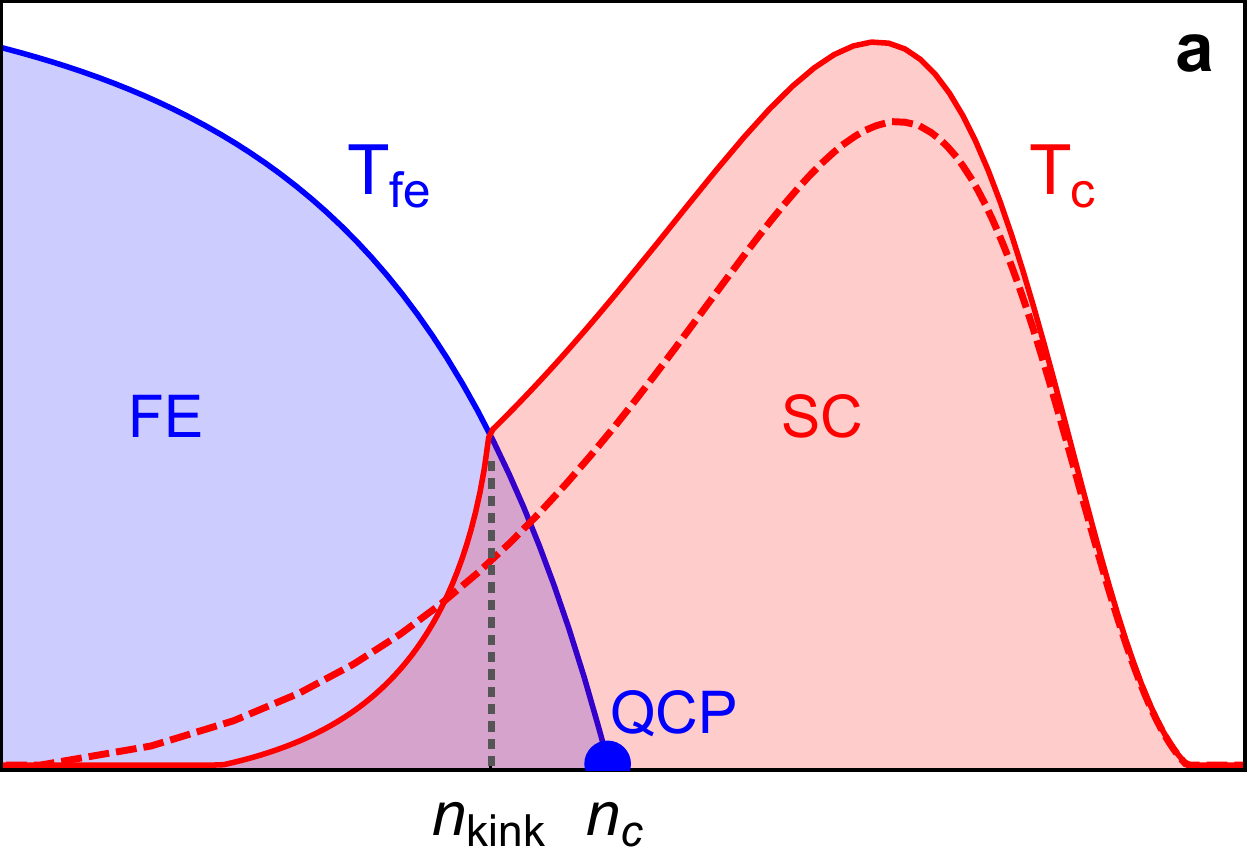}
	\includegraphics[scale=0.5]{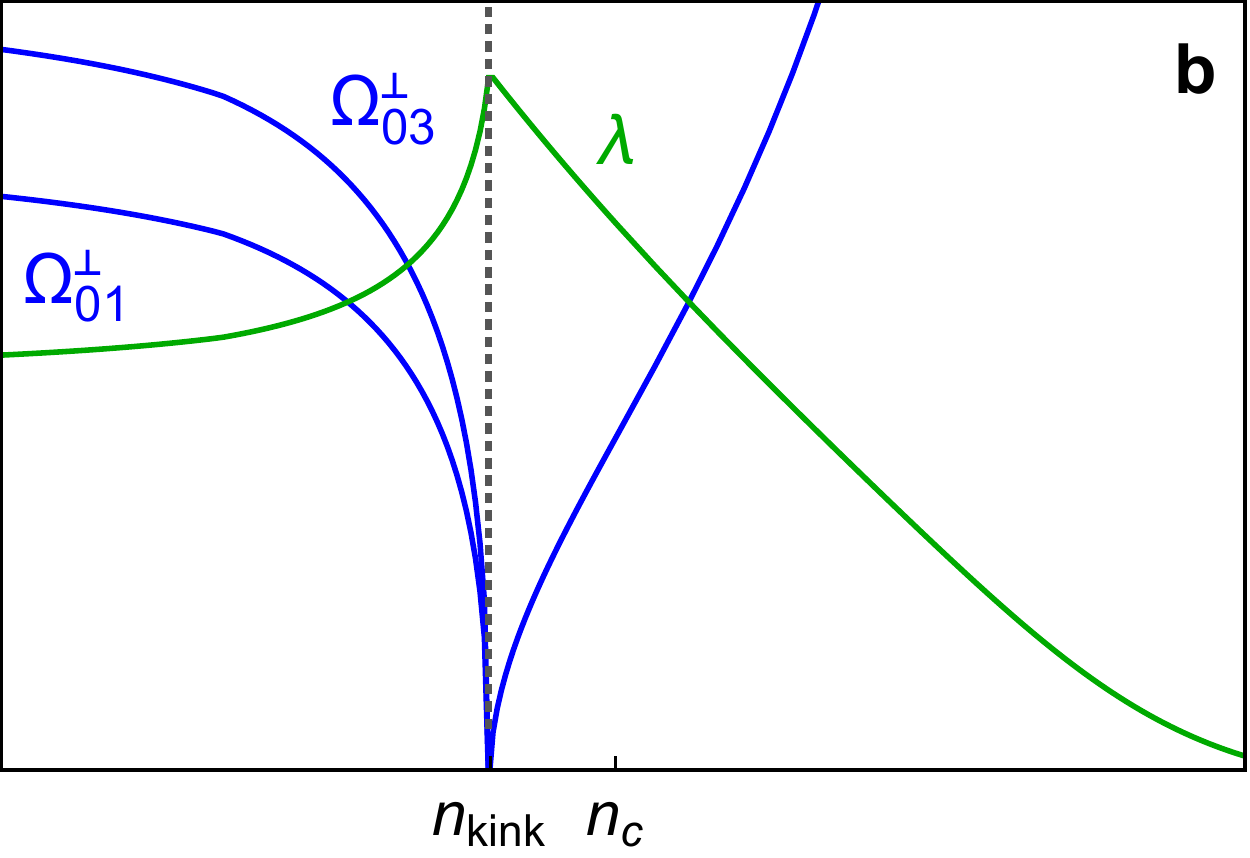}
	\caption{(a) Schematic phase diagram of our model. Red-solid and red-dashed lines are the SC transition temperatures with and without stress or cation substitution, respectively.
	(b) Doping dependence of the TO phonon energies $\Omega_{01,3}^{\perp}$  and SC coupling constant $\lambda$  along the red-solid phase
boundary $T_c$ shown in (a).}
	\label{fig:schematics}
\end{figure}
\subsection{FE phase}
The observed FE order in STO:Ca has orthorhombic symmetry~\cite{Bednorz1984a}.
Such states, however, appear as saddle points in the free energy of the 
the Hamiltonian (\ref{eq:Hamiltonian})~\cite{Cowley1980a}. 
Thus, we will assume that the FE ground state has a noncentrosymmetric tetragonal symmetry
with an order parameter $A$ along the $z$ saxis. Below $T_{fe}$, the LO frequency becomes anisotropic, while the
degenerate PE TO phonon gives rise to 
two distinct modes,
\begin{subequations} 
	\label{eq:TOT1}
	\begin{align} 
	\left( \Omega_{ {\bm q} 1 } \right)^2 &=  \left( \Omega_{0 1}^{\perp} \right)^2 + B^2 q^2, \\
	 \left( \Omega_{ {\bm q} 2 } \right)^2 &= \left( \Omega_{ {\bm q} 1 } \right)^2   +  \left[ \left( \Omega_{0 3}^{\perp} \right)^2 -  \left( \Omega_{0 1}^{\perp} \right)^2 \right]  		\left( \frac{  q_{\perp}  }{ q} \right)^2, \\ 
	 \left( \Omega_{ {\bm q} 3} \right)^2 &= \! \left( \Omega_{ {\bm q} 1 } \right)^2 \hspace{-0.13cm} +  C^2  \hspace{-0.11cm}
	 +  \left[  \left( \Omega_{0 3}^{\perp} \right)^2 \hspace{-0.16cm} - \left( \Omega_{0 1}^{\perp} \right)^2 \right]  \left( \frac{  q_z  }{ q } \right)^2,
\end{align}
\end{subequations}
\begin{figure}[htp]
	\centering
	\includegraphics[scale=0.4]{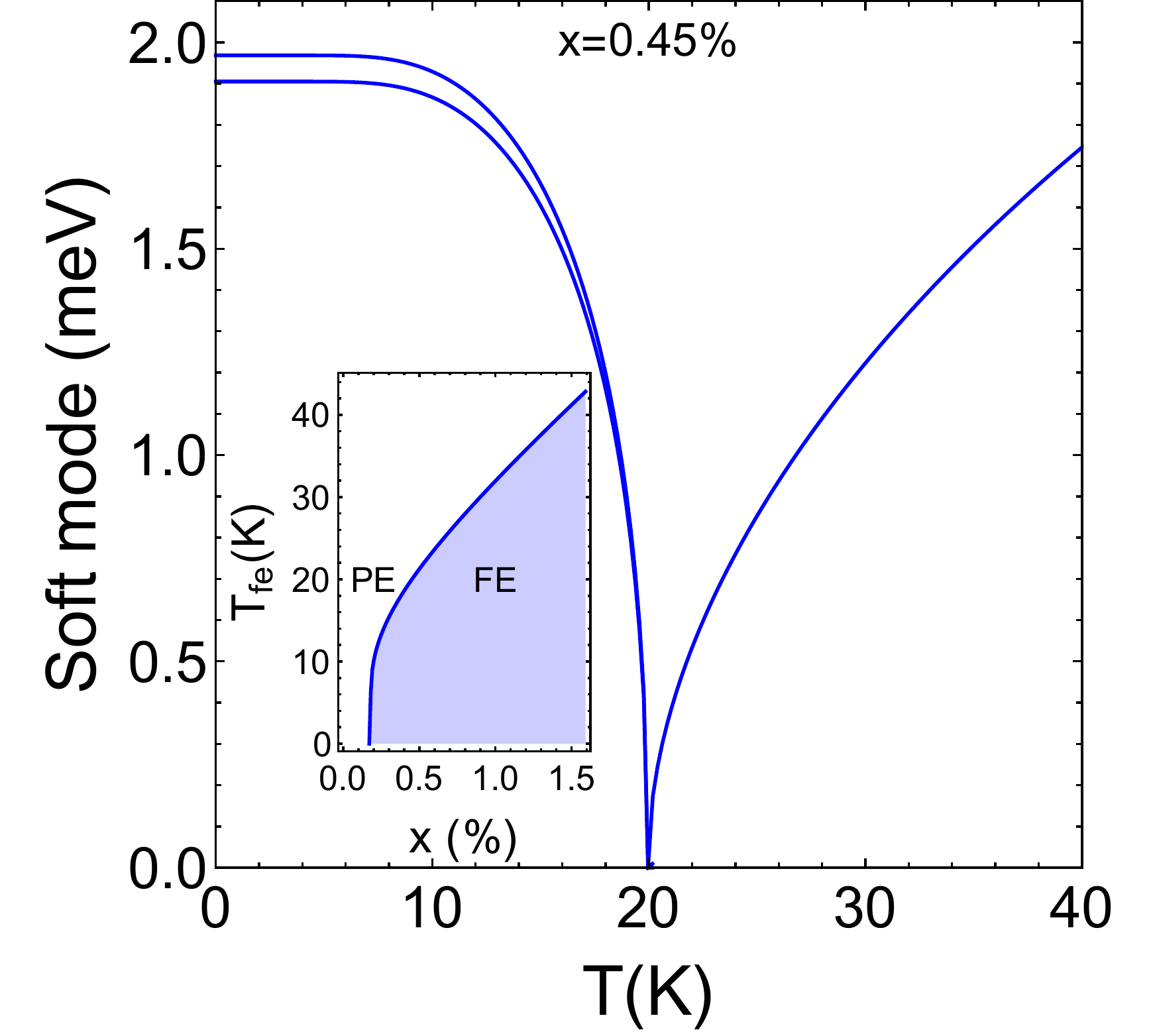}
	\caption{Calculated TO soft mode frequencies for  STO:Ca-$0.45 \%$ showing 
	 the expected split of the high-temperature TO doublet into  
	 two modes at the FE transition $T_{fe}\simeq 20\,$K. 
	Inset: Calculated phase diagram for STO:Ca-$x$.}
	\label{fig:soft_modes}
\end{figure}
where $q_{\perp}  = \sqrt{q_x^2 + q_y^2}$.
$\Omega_{0 1}^{\perp}$, and $\Omega_{0 3}^{\perp} $ are zone-center TO phonons
for ${\bm q} \perp \left(001\right)$ given as follows~\cite{Pytte1972a},
\begin{subequations}
\label{eq:TOT2}
\begin{align}
 	\left( \Omega_{0 1}^{\perp} \right)^2 &= \gamma A^2 +  \left( 2 \alpha - \gamma \right) \left( \psi_1 - \psi_3 \right) + 6 e_t \eta_t, \\
	\left( \Omega_{0 3}^{\perp} \right)^2 &= 2 \alpha A^2,\\
	 	 \left( \Omega_{03}^\perp \right)^2 &=   \omega_0^2 + 3 \alpha \left(  A^2 + \psi_3 \right) 
	 	 + 2 \left( \alpha + \gamma \right)  \psi_1 \nonumber \\ 
	 	 &~~~~~~~~~~~~~~~~~~~~~~~~~~~~~~~~~
	 	 - 2 e_a \eta_a - 4 e_t \eta_t.
\end{align}
\end{subequations}
$\psi_{1,3} = \left( 2  \omega_{1,3} \right)^{-1} \coth\left( \beta   \omega_{1,3} / 2 \right)$
are local fluctuations of polarization 
with $  \omega_{1,3} = \sqrt{ \left(\Omega_{01,3}^{\perp}\right)^2 + v_0} $, 
$\eta_a = \bra \eta_1 + \eta_2 + \eta_3  \ket =  \left( e_a / C_a \right) \left( 2 \psi_1 + A^2 + \psi_3 \right) - P /C_a$, and 
$\eta_t = \bra \eta_3 - \eta_1  \ket = \left( 3 e_t / 2 C_t \right) \left( A^2 + \psi_3 - \psi_1 \right)  $ are volume and deviatoric strains, respectively.
Equation ~(\ref{eq:TOT1}) shows that the phase
volume of critical fluctuations in the FE phase is reduced by the dipolar force.

According to  Eq.~(\ref{eq:lambda}), we therefore have,
\begin{align}
	\label{eq:lambdaFE}
	\lambda(n,T_{c}, P) = \alpha_{e-ph}^2   \sum_{\bm q} \left( \frac{1}{\Omega_{ {\bm q} 1 } } + \frac{1}{\Omega_{ {\bm q} 2 } }  + \frac{1}{\Omega_{ {\bm q} 3 } }  \right).
\end{align}
Similarly to the PE phase, the critical isotropic TO phonons make 
the largest contributions to $\lambda$.

We assume 
the same $\epsilon_F$ and $x$ dependence of the model parameters  
as that of the nonpolar phase.
Thus,
Eqs.~(\ref{eq:Tsc}) and (\ref{eq:w0})-(\ref{eq:lambdaFE})  
give the relevant transition temperatures, phonon frequencies, 
and order parameter in a self-consistent fashion.

\section{Results and Discussion}

\begin{figure*}[htp!] 
	\centering
	\includegraphics[scale=0.55]{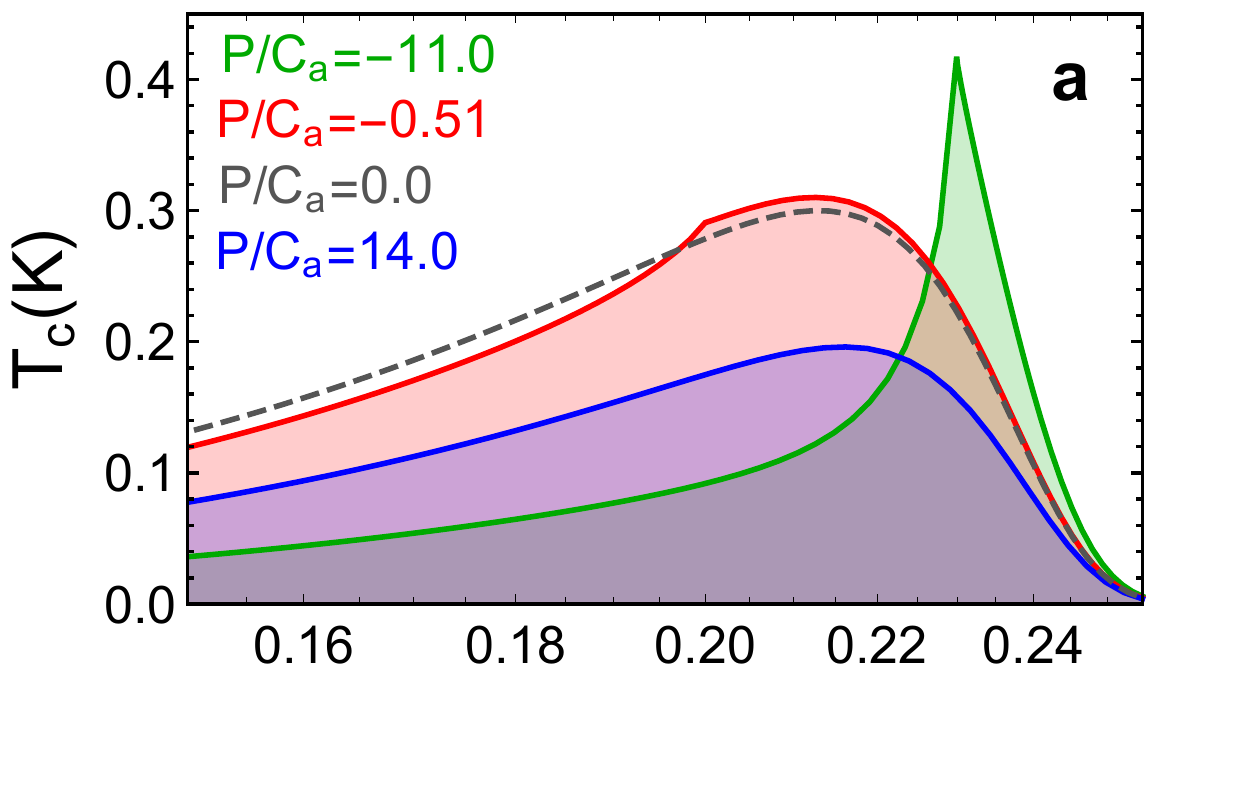} \vspace{-0.5cm}
	\includegraphics[scale=0.55]{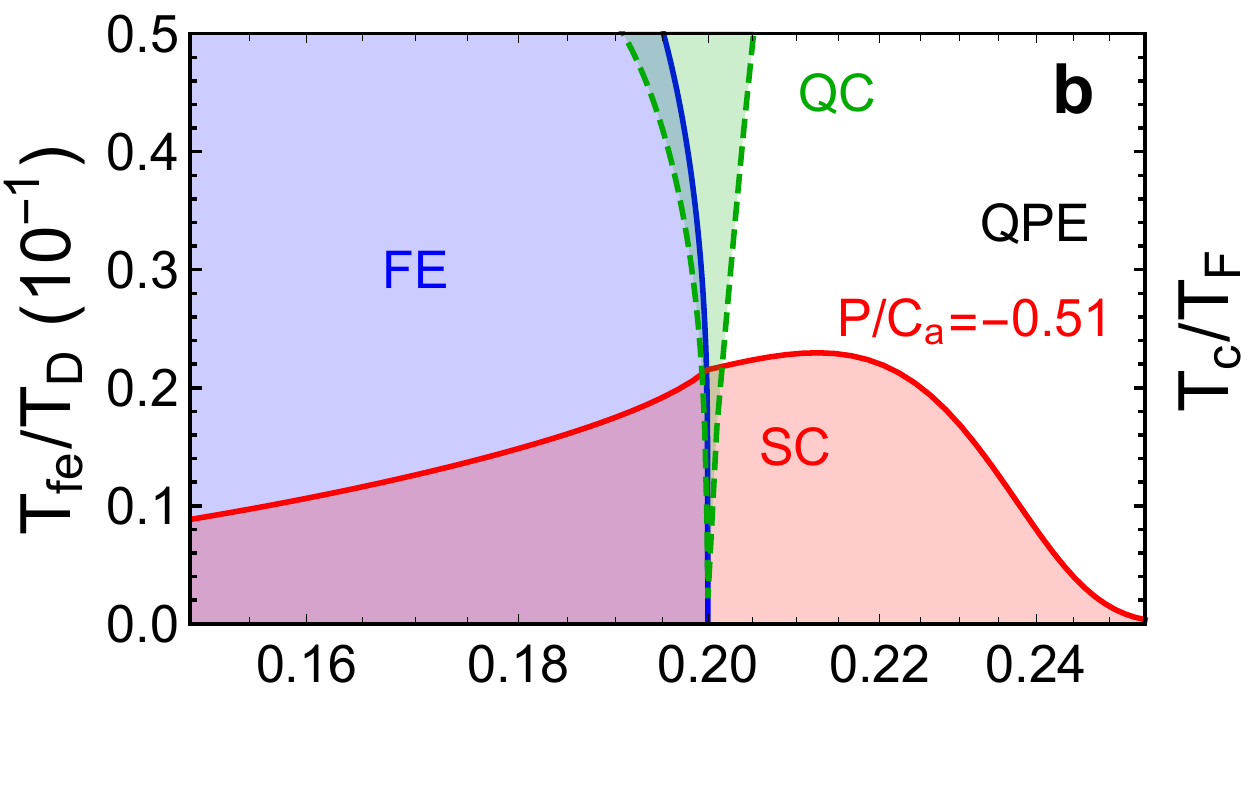}
	\includegraphics[scale=0.55]{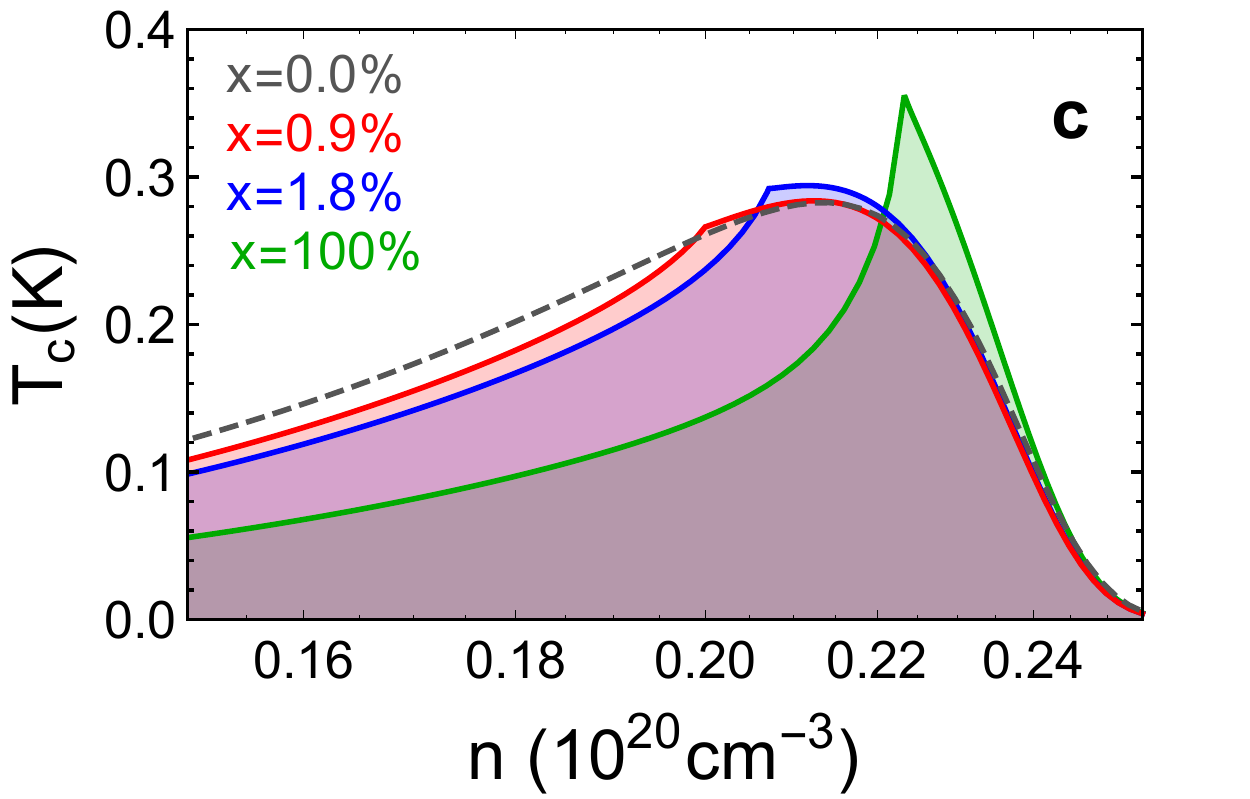}
	\includegraphics[scale=0.55]{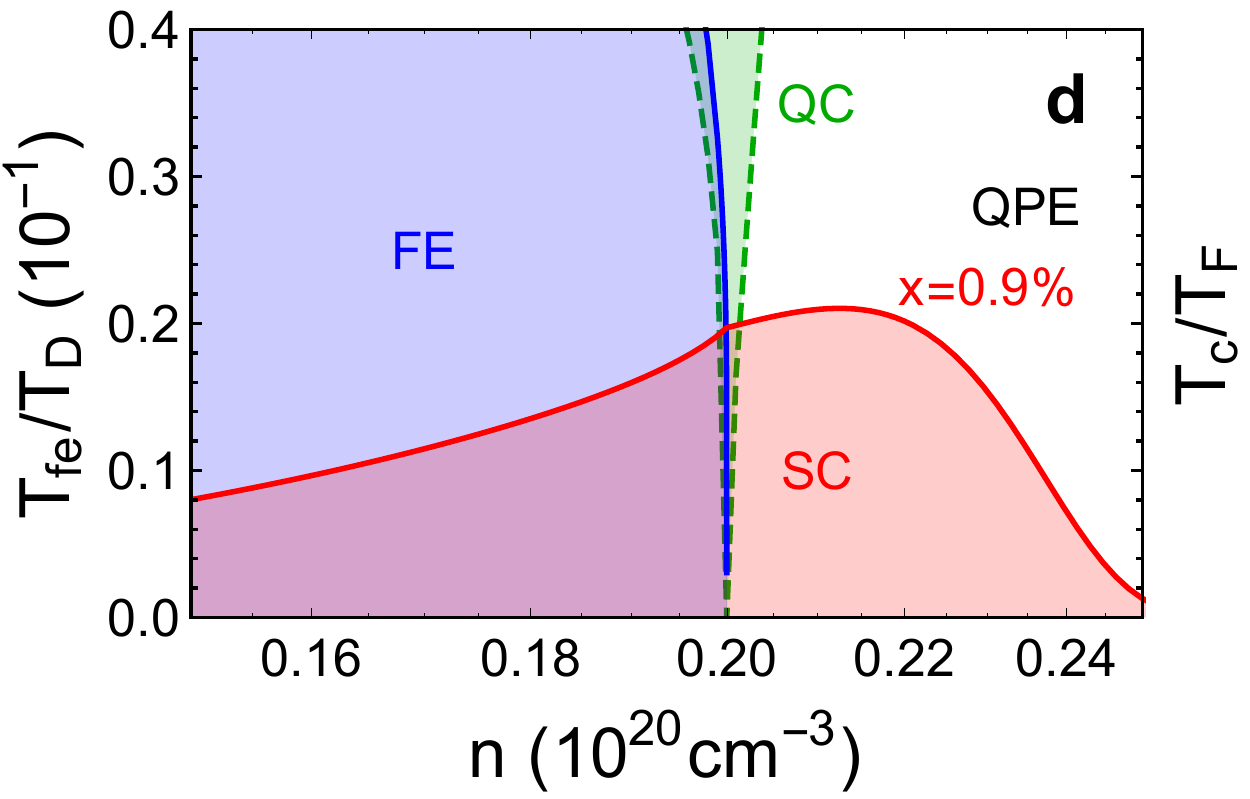}
	\caption{(a) Calculated $T_c$ for n-doped STO for several negative and positive hydrostatic
pressures.  (b) Phase diagram for n-doped STO at $P /C_a=-0.51$,
showing the overlapping FE phase (blue) and QC region (green). (c) Calculated $T_c$ for n-doped STO:Ca$-x$. (d) Phase diagram for $x=0.9\%$. }
	\label{fig:domes}
\end{figure*}

Figure~\ref{fig:schematics}~(a) shows a schematic phase diagram calculated from our model. 
In the absence of strain or cation substitution, the pairing excitations are PE at all charge densities and we find a dome-shaped $T_c$ similar to the OD models.
When long-range polar order is induced by either tuning parameter,
we find that it coexists with superconductivity
up to a charge density $n_c$
where the FE phase is terminated at the QCP.
$T_c$ is enhanced on the PE side whereas it quickly 
decreases on the FE phase.  
A kink appears at a 
charge density $n_{kink}$,
where $T_c=T_{fe}$.  
Figure ~\ref{fig:schematics}~(b) shows this is where the TO phonon condenses, which
maximizes the coupling constant, as 
as it is shown by the sharp peak in $\lambda$. 

We now apply our model to n-doped STO:Ca. 
The model parameters 
are obtained by fitting
to the observed phonon frequencies~\cite{Rischau2017a, Samara1971a, Yamada1969a}, 
FE phase diagram,~\cite{Bednorz1984a} and Debye temperature~\cite{Burns1980a} 
of STO:Ca without doping.
The elastic constants were taken from Ref.~[\onlinecite{Carpenter2007a}]. 
As it is usual, we take the continuum limit and replace the  summations in Eqs.~(\ref{eq:lambdaPE}) and~(\ref{eq:lambdaFE})
by integrals over a sphere of wave-vector cutoff radius $\Lambda$~\cite{Arce-Gamboa2017a}.
Typical results are shown in Fig.~\ref{fig:soft_modes} and reproduce well the observed behavior.
The parameters $b_1, b_2$, and $\alpha_{e-ph}$ are chosen to match the observed  QCP 
in n-doped STO:Ca-$0.9\%$ at $n_c \simeq 0.2\times 10^{20}\,$cm$^{-3}$~\cite{Rischau2017a}  
and the maximum $T_c$ ($\simeq 0.3\,$K) for n-doped STO at ambient pressure~\cite{Koonce1967a}.
The values are the following: $\omega_0=3.9i\,$meV, $\alpha=7.9\,$meV$^3$, 
$\gamma=14.8\,$meV$^3$, $B = 138.1\,$ meV \AA$^{-1}$, $C=7.7\,$meV, $ b_1= 5.0\,$ meV
$ b_2=3.4 \times 10^{-16}$,  $g=30.7$, $ \Lambda=\pi/a$, $\alpha_{e-ph}=1.5\,$meV$^{1/4}$,
$e_a=4.2\,$ meV$^2$, $e_t= 0.1\,$  meV$^2$, $C_a= 7.2 \times 10^{4}\,$ meV 
$C_t= 4.6 \times 10^{4}\,$ meV.  

We first discuss our results for applied hydrostatic pressures.
Figure~\ref{fig:domes}~(a) shows the calculated SC domes for several positive
and negative pressures. 
At zero pressure, there is a SC dome shown by the dashed line. Hydrostatic compression 
pushes the system away from the QCP by hardening  
the frequency of the TO phonons, 
therefore decreasing $\lambda$ as well as $T_c$. 
For negative pressures,  a QCP is induced within the dome and $T_c$ slightly increases on the PE side while
it decreases on the FE side except very near the kink.
By increasing negative pressure, the QCP is pushed to higher densities and the dome narrows and sharpens.
Figure~\ref{fig:domes}~(b) shows the phase diagram for doped STO under a fixed negative pressure ($P/C_a=-0.51$). 
$T_{fe}$ and $T_c$ are shown, respectively, in units of $T_D$ and $T_{F}$ to 
show that  $n_{kink} \lesssim n_c$ and that the kink lies within the
 QC region  $\Omega <  T \ll  T_D$~\cite{Rowley2014a}.  Thus, for STO,  we find that the pairing fluctuations near the kink
 are quantum critical.

We find similar results when the QCP is induced by cation substitution.
Figure~\ref{fig:domes}~(c) shows the SC domes for several Ca concentrations. Inducing  a QCP
with $x$ also generates a kink at a characteristic charge density $n_{kink}$. For $n>n_{kink}$,
$T_c$ increases, while it decreases for  $n<n_{kink}$ except very near the FE transition. 
Increasing $x$ narrows and sharpens the dome. 
Figure~\ref{fig:domes}~(d) shows the phase diagram for $x=0.9\%$. We find again that the kink occurs
nearly at the QCP and lies within the QC fan. 

We now compare our results to those of previous OD models.~\cite{Edge2015a, Kedem2016a, Dunnett2016a}. 
At the static and qualitative level considered here, our results agree with the essentials of such models in the absence of cation substitution or stress.  
However, we find that when a FE transition is induced with these tuning parameters, 
OD models predict a broadening of SC dome and a sharp peak in $T_c$. This is in stark contrast with our results. 

We now compare our results to recent experiments.
Our phase diagram is in overall qualitative and fair quantitative agreement with recent experiments 
in STO:Ca-$x$~\cite{Rischau2017a}.
While an enhancement in $T_c$ is observed near the QCP, the existing data~\cite{Rischau2017a} is too sparse to determine whether there is a kink. 
Our results are also in qualitative agreement with the observed reduction in $T_c$ with hydrostatic pressure in
n-doped STO~\cite{Rowley2018a, Pfeiffer1970a} and the very recent observation of enhanced superconductivity with tensile stress~\cite{Herrera2018a}.
Our calculated dome tends to be narrower than the observed one.~\cite{Schooley1965a, Koonce1967a, Binning1980a, Suzuki1996a} 
This is a consequence of our choice of the parametrization given in Eq.~(\ref{eq:w0}) 
for which a theory is clearly needed~\cite{Woelfe2018a}. 

\section{Conclusions}

In summary, by combining a standard model for displacive FEs with a strong-coupling theory of superconductivity, 
we have studied 
the effects of a FE quantum phase transition on the SC phase of polar semiconductors and applied it to STO.
We have shown that superconductivity is favored by 
 the FE instability from the disordered side, while the increase in
dielectric stiffness and dipolar induced anisotropies in the pairing excitations 
decrease $T_c$. A kink signature in the dome is generated by the condensation of the phonons when 
both SC and FE orders set in and when the coupling constant peaks.  
This generally occurs at charge densities below that of the QCP.
When we apply our model to doped STO, 
we find that the antiadiabatic limit 
places the kink nearly at its QCP. 
Our model is in qualitative and fairly quantitative agreement
with the recent observation of 
overlapping FE-like
and superconducting instabilities 
in n-doped STO:Ca-$x$
and the strain tuning of $T_c$  in n-doped STO.
At the qualitative and static level considered here, we find that while
these results agree with the essentials of previous work invoking OD models to describe the pairing excitations in STO, there are significant differences. 

The theoretical framework presented here and its extensions could provide insight into the intriguing role of spatial inversion symmetry breaking in two-dimensional superconductivity at 
the interface of STO-based heterostructures,~\cite{Pai2018a, *Gariglio2016a} gated KTaO$_3$~\cite{Ueno2011a},
FeSe monolayers on STO~\cite{Lee2014a},
as well as possible pairing mechanisms with
new collective excitations originated by multiferroic QCPs~\cite{Morice2017a, *Narayan2017a, *Dunnett2018b}.

Finally, to further explore the possible role of quantum structural 
transitions on electronic degrees of freedom, we speculate 
that it may be worthwhile to chemically or electrostatically dope 
other material candidates near structural QCPs such as the ionic insulators
ScF$_3$~\cite{Handunkanda2015a}~(as recently suggested~\cite{Herrera2018a}) and the mercurus halide Hg$_2$I$_2$~\cite{Occhialini2017a}.

\section{Acknowledgments}

We wish to thank Peter Littlewood and Gilbert Lonzarich for useful discussions. 
Work at the University of Costa Rica is supported by the Vice-rectory for Research
under project no. 816-B7-601, and work at Argonne National Laboratory is supported by the U.S. Department of Energy, 
Office of Basic Energy Sciences,  Materials Science Division under contract no. DE-AC02-06CH11357.
G.G.G.V. acknowledges Churchill College, the Department of Materials Science and Metallurgy and the 
Cavendish Laboratory at the University of Cambridge where part of this work was done.

%


\begin{thebibliography}{66}%
\makeatletter
\providecommand \@ifxundefined [1]{%
 \@ifx{#1\undefined}
}%
\providecommand \@ifnum [1]{%
 \ifnum #1\expandafter \@firstoftwo
 \else \expandafter \@secondoftwo
 \fi
}%
\providecommand \@ifx [1]{%
 \ifx #1\expandafter \@firstoftwo
 \else \expandafter \@secondoftwo
 \fi
}%
\providecommand \natexlab [1]{#1}%
\providecommand \enquote  [1]{``#1''}%
\providecommand \bibnamefont  [1]{#1}%
\providecommand \bibfnamefont [1]{#1}%
\providecommand \citenamefont [1]{#1}%
\providecommand \href@noop [0]{\@secondoftwo}%
\providecommand \href [0]{\begingroup \@sanitize@url \@href}%
\providecommand \@href[1]{\@@startlink{#1}\@@href}%
\providecommand \@@href[1]{\endgroup#1\@@endlink}%
\providecommand \@sanitize@url [0]{\catcode `\\12\catcode `\$12\catcode
  `\&12\catcode `\#12\catcode `\^12\catcode `\_12\catcode `\%12\relax}%
\providecommand \@@startlink[1]{}%
\providecommand \@@endlink[0]{}%
\providecommand \url  [0]{\begingroup\@sanitize@url \@url }%
\providecommand \@url [1]{\endgroup\@href {#1}{\urlprefix }}%
\providecommand \urlprefix  [0]{URL }%
\providecommand \Eprint [0]{\href }%
\providecommand \doibase [0]{http://dx.doi.org/}%
\providecommand \selectlanguage [0]{\@gobble}%
\providecommand \bibinfo  [0]{\@secondoftwo}%
\providecommand \bibfield  [0]{\@secondoftwo}%
\providecommand \translation [1]{[#1]}%
\providecommand \BibitemOpen [0]{}%
\providecommand \bibitemStop [0]{}%
\providecommand \bibitemNoStop [0]{.\EOS\space}%
\providecommand \EOS [0]{\spacefactor3000\relax}%
\providecommand \BibitemShut  [1]{\csname bibitem#1\endcsname}%
\let\auto@bib@innerbib\@empty
\bibitem [{\citenamefont {Schooley}\ \emph {et~al.}(1964)\citenamefont
  {Schooley}, \citenamefont {Hosler},\ and\ \citenamefont
  {Cohen}}]{Schooley1964a}%
  \BibitemOpen
  \bibfield  {author} {\bibinfo {author} {\bibfnamefont {J.~F.}\ \bibnamefont
  {Schooley}}, \bibinfo {author} {\bibfnamefont {W.~R.}\ \bibnamefont
  {Hosler}}, \ and\ \bibinfo {author} {\bibfnamefont {M.~L.}\ \bibnamefont
  {Cohen}},\ }\bibfield  {title} {\enquote {\bibinfo {title}
  {{Superconductivity in semiconducting SrTiO$_3$}},}\ }\href
  {https://link.aps.org/doi/10.1103/PhysRevLett.12.474} {\bibfield  {journal}
  {\bibinfo  {journal} {Phys. Rev. Lett.}\ }\textbf {\bibinfo {volume} {12}},\
  \bibinfo {pages} {474} (\bibinfo {year} {1964})}\BibitemShut {NoStop}%
\bibitem [{\citenamefont {Schooley}\ \emph {et~al.}(1965)\citenamefont
  {Schooley}, \citenamefont {Hosler}, \citenamefont {Ambler}, \citenamefont
  {Becker}, \citenamefont {Cohen},\ and\ \citenamefont
  {Koonce}}]{Schooley1965a}%
  \BibitemOpen
  \bibfield  {author} {\bibinfo {author} {\bibfnamefont {J.~F.}\ \bibnamefont
  {Schooley}}, \bibinfo {author} {\bibfnamefont {W.~R.}\ \bibnamefont
  {Hosler}}, \bibinfo {author} {\bibfnamefont {E.}~\bibnamefont {Ambler}},
  \bibinfo {author} {\bibfnamefont {J.~H.}\ \bibnamefont {Becker}}, \bibinfo
  {author} {\bibfnamefont {M.~L.}\ \bibnamefont {Cohen}}, \ and\ \bibinfo
  {author} {\bibfnamefont {C.~S.}\ \bibnamefont {Koonce}},\ }\bibfield  {title}
  {\enquote {\bibinfo {title} {{Dependence of the superconducting transition
  temperature on carrier concentration in semiconducting
  SrTi${\mathrm{O}}_{3}$}},}\ }\href {\doibase 10.1103/PhysRevLett.14.305}
  {\bibfield  {journal} {\bibinfo  {journal} {Phys. Rev. Lett.}\ }\textbf
  {\bibinfo {volume} {14}},\ \bibinfo {pages} {305} (\bibinfo {year}
  {1965})}\BibitemShut {NoStop}%
\bibitem [{\citenamefont {Koonce}\ \emph {et~al.}(1967)\citenamefont {Koonce},
  \citenamefont {Cohen}, \citenamefont {Schooley}, \citenamefont {Hosler},\
  and\ \citenamefont {Pfeiffer}}]{Koonce1967a}%
  \BibitemOpen
  \bibfield  {author} {\bibinfo {author} {\bibfnamefont {C.~S.}\ \bibnamefont
  {Koonce}}, \bibinfo {author} {\bibfnamefont {M.~L.}\ \bibnamefont {Cohen}},
  \bibinfo {author} {\bibfnamefont {J.~F.}\ \bibnamefont {Schooley}}, \bibinfo
  {author} {\bibfnamefont {W.~R.}\ \bibnamefont {Hosler}}, \ and\ \bibinfo
  {author} {\bibfnamefont {E.~R.}\ \bibnamefont {Pfeiffer}},\ }\bibfield
  {title} {\enquote {\bibinfo {title} {{Superconducting transition temperatures
  of semiconducting SrTiO$_{3}$}},}\ }\href {\doibase 10.1103/PhysRev.163.380}
  {\bibfield  {journal} {\bibinfo  {journal} {Phys. Rev.}\ }\textbf {\bibinfo
  {volume} {163}},\ \bibinfo {pages} {380} (\bibinfo {year}
  {1967})}\BibitemShut {NoStop}%
\bibitem [{\citenamefont {Binnig}\ \emph {et~al.}(1980)\citenamefont {Binnig},
  \citenamefont {Baratoff}, \citenamefont {Hoenig},\ and\ \citenamefont
  {Bednorz}}]{Binning1980a}%
  \BibitemOpen
  \bibfield  {author} {\bibinfo {author} {\bibfnamefont {G.}~\bibnamefont
  {Binnig}}, \bibinfo {author} {\bibfnamefont {A.}~\bibnamefont {Baratoff}},
  \bibinfo {author} {\bibfnamefont {H.~E.}\ \bibnamefont {Hoenig}}, \ and\
  \bibinfo {author} {\bibfnamefont {J.~G.}\ \bibnamefont {Bednorz}},\
  }\bibfield  {title} {\enquote {\bibinfo {title} {{Two-band superconductivity
  in Nb-doped SrTi${\mathrm{O}}_{3}$}},}\ }\href
  {https://link.aps.org/doi/10.1103/PhysRevLett.45.1352} {\bibfield  {journal}
  {\bibinfo  {journal} {Phys. Rev. Lett.}\ }\textbf {\bibinfo {volume} {45}},\
  \bibinfo {pages} {1352} (\bibinfo {year} {1980})}\BibitemShut {NoStop}%
\bibitem [{\citenamefont {Suzuki}\ \emph {et~al.}(1996)\citenamefont {Suzuki},
  \citenamefont {Bando}, \citenamefont {Ootuka}, \citenamefont {Inoue},
  \citenamefont {Yamamoto}, \citenamefont {Takahashi},\ and\ \citenamefont
  {Nishihara}}]{Suzuki1996a}%
  \BibitemOpen
  \bibfield  {author} {\bibinfo {author} {\bibfnamefont {Hiroshi}\ \bibnamefont
  {Suzuki}}, \bibinfo {author} {\bibfnamefont {Hiroshi}\ \bibnamefont {Bando}},
  \bibinfo {author} {\bibfnamefont {Youiti}\ \bibnamefont {Ootuka}}, \bibinfo
  {author} {\bibfnamefont {Isao~H.}\ \bibnamefont {Inoue}}, \bibinfo {author}
  {\bibfnamefont {Tetsuya}\ \bibnamefont {Yamamoto}}, \bibinfo {author}
  {\bibfnamefont {Kazuhiko}\ \bibnamefont {Takahashi}}, \ and\ \bibinfo
  {author} {\bibfnamefont {Yoshikazu}\ \bibnamefont {Nishihara}},\ }\bibfield
  {title} {\enquote {\bibinfo {title} {{Superconductivity in single-crystalline
  Sr$_{1-x}$La$_x$TiO$_3$}},}\ }\href {\doibase 10.1143/JPSJ.65.1529}
  {\bibfield  {journal} {\bibinfo  {journal} {J. Phys. Soc. Jpn.}\ }\textbf
  {\bibinfo {volume} {65}},\ \bibinfo {pages} {1529} (\bibinfo {year}
  {1996})}\BibitemShut {NoStop}%
\bibitem [{\citenamefont {Eagles}(1986)}]{Eagles1986a}%
  \BibitemOpen
  \bibfield  {author} {\bibinfo {author} {\bibfnamefont {D.~M.}\ \bibnamefont
  {Eagles}},\ }\bibfield  {title} {\enquote {\bibinfo {title}
  {{Superconductivity at very low carrier concentrations and indications of a
  charged bose gas in SrTi$_{0.97} $Zr$_{0.03}$O$_3$}},}\ }\href
  {http://www.sciencedirect.com/science/article/pii/0038109886907301}
  {\bibfield  {journal} {\bibinfo  {journal} {Solid State Comm.}\ }\textbf
  {\bibinfo {volume} {60}},\ \bibinfo {pages} {521} (\bibinfo {year}
  {1986})}\BibitemShut {NoStop}%
\bibitem [{\citenamefont {Eagles}\ \emph {et~al.}(1989)\citenamefont {Eagles},
  \citenamefont {Tainsh},\ and\ \citenamefont {Andrikidis}}]{Eagles1989a}%
  \BibitemOpen
  \bibfield  {author} {\bibinfo {author} {\bibfnamefont {D.~M.}\ \bibnamefont
  {Eagles}}, \bibinfo {author} {\bibfnamefont {R.~J.}\ \bibnamefont {Tainsh}},
  \ and\ \bibinfo {author} {\bibfnamefont {C.}~\bibnamefont {Andrikidis}},\
  }\bibfield  {title} {\enquote {\bibinfo {title} {{Evidence for pairing
  without superconductivity from resistance between 130 mK and 70 mK in a
  specimen of ceramic Zr-doped SrTiO$_3$}},}\ }\href
  {http://www.sciencedirect.com/science/article/pii/0921453489904668}
  {\bibfield  {journal} {\bibinfo  {journal} {Physica C}\ }\textbf {\bibinfo
  {volume} {157}},\ \bibinfo {pages} {48} (\bibinfo {year} {1989})}\BibitemShut
  {NoStop}%
\bibitem [{\citenamefont {Tainsh}\ and\ \citenamefont
  {Andrikidis}(1986)}]{Tainsh1986a}%
  \BibitemOpen
  \bibfield  {author} {\bibinfo {author} {\bibfnamefont {R.~J.}\ \bibnamefont
  {Tainsh}}\ and\ \bibinfo {author} {\bibfnamefont {C.}~\bibnamefont
  {Andrikidis}},\ }\bibfield  {title} {\enquote {\bibinfo {title}
  {{Superconducting transitions from states with low normal conductivity in
  ceramic SrTi$_{0.97}$Zr$_{0.03}$O$_3$}},}\ }\href {\doibase
  https://doi.org/10.1016/0038-1098(86)90729-5} {\bibfield  {journal} {\bibinfo
   {journal} {Solid State Comm.}\ }\textbf {\bibinfo {volume} {60}},\ \bibinfo
  {pages} {517} (\bibinfo {year} {1986})}\BibitemShut {NoStop}%
\bibitem [{\citenamefont {Eagles}(2016)}]{Eagles2016a}%
  \BibitemOpen
  \bibfield  {author} {\bibinfo {author} {\bibfnamefont {D.~M.}\ \bibnamefont
  {Eagles}},\ }\bibfield  {title} {\enquote {\bibinfo {title} {{Comment on two
  papers claiming records for the lowest carrier concentration at which
  superconductivity has been observed}},}\ }\href
  {https://arxiv.org/abs/1604.05660} {\bibfield  {journal} {\bibinfo  {journal}
  {arXiv:1604.05660}\ } (\bibinfo {year} {2016})}\BibitemShut {NoStop}%
\bibitem [{\citenamefont {Lin}\ \emph {et~al.}(2013)\citenamefont {Lin},
  \citenamefont {Zhu}, \citenamefont {Fauqu\'e},\ and\ \citenamefont
  {Behnia}}]{Lin2013a}%
  \BibitemOpen
  \bibfield  {author} {\bibinfo {author} {\bibfnamefont {X.}~\bibnamefont
  {Lin}}, \bibinfo {author} {\bibfnamefont {Z.}~\bibnamefont {Zhu}}, \bibinfo
  {author} {\bibfnamefont {B.}~\bibnamefont {Fauqu\'e}}, \ and\ \bibinfo
  {author} {\bibfnamefont {K.}~\bibnamefont {Behnia}},\ }\bibfield  {title}
  {\enquote {\bibinfo {title} {Fermi surface of the most dilute
  superconductor},}\ }\href {\doibase 10.1103/PhysRevX.3.021002} {\bibfield
  {journal} {\bibinfo  {journal} {Phys. Rev. X}\ }\textbf {\bibinfo {volume}
  {3}},\ \bibinfo {pages} {021002} (\bibinfo {year} {2013})}\BibitemShut
  {NoStop}%
\bibitem [{\citenamefont {Carbotte}(1990)}]{Cabrotte1990a}%
  \BibitemOpen
  \bibfield  {author} {\bibinfo {author} {\bibfnamefont {J.~P.}\ \bibnamefont
  {Carbotte}},\ }\bibfield  {title} {\enquote {\bibinfo {title} {Properties of
  boson-exchange superconductors},}\ }\href
  {https://link.aps.org/doi/10.1103/RevModPhys.62.1027} {\bibfield  {journal}
  {\bibinfo  {journal} {Rev. Mod. Phys.}\ }\textbf {\bibinfo {volume} {62}},\
  \bibinfo {pages} {1027} (\bibinfo {year} {1990})}\BibitemShut {NoStop}%
\bibitem [{\citenamefont {Burns}(1980)}]{Burns1980a}%
  \BibitemOpen
  \bibfield  {author} {\bibinfo {author} {\bibfnamefont {G.}~\bibnamefont
  {Burns}},\ }\bibfield  {title} {\enquote {\bibinfo {title} {Comment on the
  low temperature specific heat of ferroelectrics, antiferroelectrics, and
  related materials},}\ }\href
  {http://www.sciencedirect.com/science/article/pii/0038109880910297}
  {\bibfield  {journal} {\bibinfo  {journal} {Solid State Comm.}\ }\textbf
  {\bibinfo {volume} {35}},\ \bibinfo {pages} {811} (\bibinfo {year}
  {1980})}\BibitemShut {NoStop}%
\bibitem [{\citenamefont {Gor'kov}(2016)}]{Gorkov2016a}%
  \BibitemOpen
  \bibfield  {author} {\bibinfo {author} {\bibfnamefont {L.~P.}\ \bibnamefont
  {Gor'kov}},\ }\bibfield  {title} {\enquote {\bibinfo {title} {{Phonon
  mechanism in the most dilute superconductor n-type SrTiO$_3$}},}\ }\href
  {\doibase 10.1073/pnas.1604145113} {\bibfield  {journal} {\bibinfo  {journal}
  {Proc. Natl. Acad. Sci.}\ }\textbf {\bibinfo {volume} {113}},\ \bibinfo
  {pages} {4646} (\bibinfo {year} {2016})}\BibitemShut {NoStop}%
\bibitem [{\citenamefont {Ruhman}\ and\ \citenamefont
  {Lee}(2016)}]{Ruhman2016a}%
  \BibitemOpen
  \bibfield  {author} {\bibinfo {author} {\bibfnamefont {J.}~\bibnamefont
  {Ruhman}}\ and\ \bibinfo {author} {\bibfnamefont {P.~A.}\ \bibnamefont
  {Lee}},\ }\bibfield  {title} {\enquote {\bibinfo {title} {Superconductivity
  at very low density: the case of strontium titanate},}\ }\href
  {https://link.aps.org/doi/10.1103/PhysRevB.94.224515} {\bibfield  {journal}
  {\bibinfo  {journal} {Phys. Rev. B}\ }\textbf {\bibinfo {volume} {94}},\
  \bibinfo {pages} {224515} (\bibinfo {year} {2016})}\BibitemShut {NoStop}%
\bibitem [{\citenamefont {Takada}(1980)}]{Takada1980a}%
  \BibitemOpen
  \bibfield  {author} {\bibinfo {author} {\bibfnamefont {Y.}~\bibnamefont
  {Takada}},\ }\bibfield  {title} {\enquote {\bibinfo {title} {{Theory of
  superconductivity in polar semiconductors and its application to n-type
  semiconducting SrTiO$_3$}},}\ }\href {\doibase 10.1143/JPSJ.49.1267}
  {\bibfield  {journal} {\bibinfo  {journal} {J. Phys. Soc. Jpn.}\ }\textbf
  {\bibinfo {volume} {49}},\ \bibinfo {pages} {1267--1275} (\bibinfo {year}
  {1980})}\BibitemShut {NoStop}%
\bibitem [{\citenamefont {Edelman}\ and\ \citenamefont
  {Littlewood}(2017)}]{Edelman2017a}%
  \BibitemOpen
  \bibfield  {author} {\bibinfo {author} {\bibfnamefont {A.}~\bibnamefont
  {Edelman}}\ and\ \bibinfo {author} {\bibfnamefont {P.~B.}\ \bibnamefont
  {Littlewood}},\ }\bibfield  {title} {\enquote {\bibinfo {title} {A41.00011 :
  Polaron-plasmon superconductivity in strontium titanate},}\ }\href
  {http://adsabs.harvard.edu/abs/2017APS..MARA41011E} {\bibfield  {journal}
  {\bibinfo  {journal} {APS March Meeting}\ } (\bibinfo {year}
  {2017})}\BibitemShut {NoStop}%
\bibitem [{\citenamefont {Swartz}\ \emph {et~al.}(2018)\citenamefont {Swartz},
  \citenamefont {Inoue}, \citenamefont {Merz}, \citenamefont {Hikita},
  \citenamefont {Raghu}, \citenamefont {Devereaux}, \citenamefont {Johnston},\
  and\ \citenamefont {Hwang}}]{Swartz2018a}%
  \BibitemOpen
  \bibfield  {author} {\bibinfo {author} {\bibfnamefont {A.~G.}\ \bibnamefont
  {Swartz}}, \bibinfo {author} {\bibfnamefont {H.}~\bibnamefont {Inoue}},
  \bibinfo {author} {\bibfnamefont {T.~A.}\ \bibnamefont {Merz}}, \bibinfo
  {author} {\bibfnamefont {Y.}~\bibnamefont {Hikita}}, \bibinfo {author}
  {\bibfnamefont {S.}~\bibnamefont {Raghu}}, \bibinfo {author} {\bibfnamefont
  {T.~P.}\ \bibnamefont {Devereaux}}, \bibinfo {author} {\bibfnamefont
  {S.}~\bibnamefont {Johnston}}, \ and\ \bibinfo {author} {\bibfnamefont
  {H.~Y.}\ \bibnamefont {Hwang}},\ }\bibfield  {title} {\enquote {\bibinfo
  {title} {Polaronic behavior in a weak-coupling superconductor},}\ }\href
  {\doibase 10.1073/pnas.1713916115} {\bibfield  {journal} {\bibinfo  {journal}
  {Proc. Nat. Acad. Sci.}\ }\textbf {\bibinfo {volume} {115}},\ \bibinfo
  {pages} {1475} (\bibinfo {year} {2018})}\BibitemShut {NoStop}%
\bibitem [{\citenamefont {Collignon}\ \emph {et~al.}(2018)\citenamefont
  {Collignon}, \citenamefont {Lin}, \citenamefont {Rischau}, \citenamefont
  {Fauqu{\'e}},\ and\ \citenamefont {Behnia}}]{Collignon2018a}%
  \BibitemOpen
  \bibfield  {author} {\bibinfo {author} {\bibfnamefont {C.}~\bibnamefont
  {Collignon}}, \bibinfo {author} {\bibfnamefont {X.}~\bibnamefont {Lin}},
  \bibinfo {author} {\bibfnamefont {C.~W.}\ \bibnamefont {Rischau}}, \bibinfo
  {author} {\bibfnamefont {B.}~\bibnamefont {Fauqu{\'e}}}, \ and\ \bibinfo
  {author} {\bibfnamefont {K.}~\bibnamefont {Behnia}},\ }\bibfield  {title}
  {\enquote {\bibinfo {title} {{Metallicity and superconductivity in doped
  strontium titanate}},}\ }\href {https://arxiv.org/abs/1804.07067} {\bibfield
  {journal} {\bibinfo  {journal} {arXiv:1804.07067}\ } (\bibinfo {year}
  {2018})}\BibitemShut {NoStop}%
\bibitem [{\citenamefont {Rowley}\ \emph {et~al.}(2014)\citenamefont {Rowley},
  \citenamefont {Spalek}, \citenamefont {Smith}, \citenamefont {Dean},
  \citenamefont {Itoh}, \citenamefont {Scott}, \citenamefont {Lonzarich},\ and\
  \citenamefont {Saxena}}]{Rowley2014a}%
  \BibitemOpen
  \bibfield  {author} {\bibinfo {author} {\bibfnamefont {S.~E.}\ \bibnamefont
  {Rowley}}, \bibinfo {author} {\bibfnamefont {L.~J.}\ \bibnamefont {Spalek}},
  \bibinfo {author} {\bibfnamefont {R.~P.}\ \bibnamefont {Smith}}, \bibinfo
  {author} {\bibfnamefont {M.~P.~M.}\ \bibnamefont {Dean}}, \bibinfo {author}
  {\bibfnamefont {M.}~\bibnamefont {Itoh}}, \bibinfo {author} {\bibfnamefont
  {J.~F.}\ \bibnamefont {Scott}}, \bibinfo {author} {\bibfnamefont {G.~G.}\
  \bibnamefont {Lonzarich}}, \ and\ \bibinfo {author} {\bibfnamefont {S.~S.}\
  \bibnamefont {Saxena}},\ }\bibfield  {title} {\enquote {\bibinfo {title}
  {{Ferroelectric quantum criticality}},}\ }\href
  {https://www.nature.com/articles/nphys2924} {\bibfield  {journal} {\bibinfo
  {journal} {Nat. Phys.}\ }\textbf {\bibinfo {volume} {10}},\ \bibinfo {pages}
  {367} (\bibinfo {year} {2014})}\BibitemShut {NoStop}%
\bibitem [{\citenamefont {Edge}\ \emph {et~al.}(2015)\citenamefont {Edge},
  \citenamefont {Kedem}, \citenamefont {Aschauer}, \citenamefont {Spaldin},\
  and\ \citenamefont {Balatsky}}]{Edge2015a}%
  \BibitemOpen
  \bibfield  {author} {\bibinfo {author} {\bibfnamefont {J.~M.}\ \bibnamefont
  {Edge}}, \bibinfo {author} {\bibfnamefont {Y.}~\bibnamefont {Kedem}},
  \bibinfo {author} {\bibfnamefont {U.}~\bibnamefont {Aschauer}}, \bibinfo
  {author} {\bibfnamefont {N.~A.}\ \bibnamefont {Spaldin}}, \ and\ \bibinfo
  {author} {\bibfnamefont {A.~V.}\ \bibnamefont {Balatsky}},\ }\bibfield
  {title} {\enquote {\bibinfo {title} {{Quantum critical origin of the
  superconducting dome in SrTiO$_3$}},}\ }\href
  {https://link.aps.org/doi/10.1103/PhysRevLett.115.247002} {\bibfield
  {journal} {\bibinfo  {journal} {Phys. Rev. Lett.}\ }\textbf {\bibinfo
  {volume} {115}},\ \bibinfo {pages} {247002} (\bibinfo {year}
  {2015})}\BibitemShut {NoStop}%
\bibitem [{\citenamefont {Rischau}\ \emph {et~al.}(2017)\citenamefont
  {Rischau}, \citenamefont {Lin}, \citenamefont {Grams}, \citenamefont {Finck},
  \citenamefont {Harms}, \citenamefont {Engelmayer}, \citenamefont {Lorenz},
  \citenamefont {Gallais}, \citenamefont {Fauque}, \citenamefont {Hemberger},\
  and\ \citenamefont {Behnia}}]{Rischau2017a}%
  \BibitemOpen
  \bibfield  {author} {\bibinfo {author} {\bibfnamefont {C.~W.}\ \bibnamefont
  {Rischau}}, \bibinfo {author} {\bibfnamefont {X.}~\bibnamefont {Lin}},
  \bibinfo {author} {\bibfnamefont {C.~P.}\ \bibnamefont {Grams}}, \bibinfo
  {author} {\bibfnamefont {D.}~\bibnamefont {Finck}}, \bibinfo {author}
  {\bibfnamefont {S.}~\bibnamefont {Harms}}, \bibinfo {author} {\bibfnamefont
  {J.}~\bibnamefont {Engelmayer}}, \bibinfo {author} {\bibfnamefont
  {T.}~\bibnamefont {Lorenz}}, \bibinfo {author} {\bibfnamefont
  {Y.}~\bibnamefont {Gallais}}, \bibinfo {author} {\bibfnamefont
  {B.}~\bibnamefont {Fauque}}, \bibinfo {author} {\bibfnamefont
  {J.}~\bibnamefont {Hemberger}}, \ and\ \bibinfo {author} {\bibfnamefont
  {K.}~\bibnamefont {Behnia}},\ }\bibfield  {title} {\enquote {\bibinfo {title}
  {{A ferroelectric quantum phase transition inside the superconducting dome of
  Sr$_{1-x}$Ca$_x$TiO$_{3 - \delta}$}},}\ }\href
  {http://dx.doi.org/10.1038/nphys4085} {\bibfield  {journal} {\bibinfo
  {journal} {Nat. Phys.}\ }\textbf {\bibinfo {volume} {13}},\ \bibinfo {pages}
  {643} (\bibinfo {year} {2017})}\BibitemShut {NoStop}%
\bibitem [{\citenamefont {Stucky}\ \emph {et~al.}(2016)\citenamefont {Stucky},
  \citenamefont {Scheerer}, \citenamefont {Ren}, \citenamefont {Jaccard},
  \citenamefont {Poumirol}, \citenamefont {Barreteau}, \citenamefont
  {Giannini},\ and\ \citenamefont {van~der Marel}}]{Stucky2016a}%
  \BibitemOpen
  \bibfield  {author} {\bibinfo {author} {\bibfnamefont {A.}~\bibnamefont
  {Stucky}}, \bibinfo {author} {\bibfnamefont {G.~W.}\ \bibnamefont
  {Scheerer}}, \bibinfo {author} {\bibfnamefont {Z.}~\bibnamefont {Ren}},
  \bibinfo {author} {\bibfnamefont {D.}~\bibnamefont {Jaccard}}, \bibinfo
  {author} {\bibfnamefont {J.-M.}\ \bibnamefont {Poumirol}}, \bibinfo {author}
  {\bibfnamefont {C.}~\bibnamefont {Barreteau}}, \bibinfo {author}
  {\bibfnamefont {E.}~\bibnamefont {Giannini}}, \ and\ \bibinfo {author}
  {\bibfnamefont {D.}~\bibnamefont {van~der Marel}},\ }\bibfield  {title}
  {\enquote {\bibinfo {title} {{Isotope effect in superconducting n-doped
  SrTiO$_3$}},}\ }\href {http://dx.doi.org/10.1038/srep37582} {\bibfield
  {journal} {\bibinfo  {journal} {Sci. Rep.}\ }\textbf {\bibinfo {volume}
  {6}},\ \bibinfo {pages} {37582} (\bibinfo {year} {2016})}\BibitemShut
  {NoStop}%
\bibitem [{\citenamefont {Herrera}\ \emph {et~al.}(2018)\citenamefont
  {Herrera}, \citenamefont {Cerbin}, \citenamefont {Dunnett}, \citenamefont
  {Balatsky},\ and\ \citenamefont {Sochnikov}}]{Herrera2018a}%
  \BibitemOpen
  \bibfield  {author} {\bibinfo {author} {\bibfnamefont {C.}~\bibnamefont
  {Herrera}}, \bibinfo {author} {\bibfnamefont {J.}~\bibnamefont {Cerbin}},
  \bibinfo {author} {\bibfnamefont {K.}~\bibnamefont {Dunnett}}, \bibinfo
  {author} {\bibfnamefont {A.~V.}\ \bibnamefont {Balatsky}}, \ and\ \bibinfo
  {author} {\bibfnamefont {I.}~\bibnamefont {Sochnikov}},\ }\bibfield  {title}
  {\enquote {\bibinfo {title} {{Strain-engineered interaction of quantum polar
  and superconducting phases}},}\ }\href {https://arxiv.org/abs/1808.03739}
  {\bibfield  {journal} {\bibinfo  {journal} {arXiv:1808.03739}\ } (\bibinfo
  {year} {2018})}\BibitemShut {NoStop}%
\bibitem [{\citenamefont {M\"uller}\ and\ \citenamefont
  {Burkard}(1979)}]{Muller1979a}%
  \BibitemOpen
  \bibfield  {author} {\bibinfo {author} {\bibfnamefont {K.~A.}\ \bibnamefont
  {M\"uller}}\ and\ \bibinfo {author} {\bibfnamefont {H.}~\bibnamefont
  {Burkard}},\ }\bibfield  {title} {\enquote {\bibinfo {title}
  {{SrTi${\mathrm{O}}_{3}$: An intrinsic quantum paraelectric below 4 K}},}\
  }\href {\doibase 10.1103/PhysRevB.19.3593} {\bibfield  {journal} {\bibinfo
  {journal} {Phys. Rev. B}\ }\textbf {\bibinfo {volume} {19}},\ \bibinfo
  {pages} {3593} (\bibinfo {year} {1979})}\BibitemShut {NoStop}%
\bibitem [{\citenamefont {Yamada}\ and\ \citenamefont
  {Shirane}(1969)}]{Yamada1969a}%
  \BibitemOpen
  \bibfield  {author} {\bibinfo {author} {\bibfnamefont {Y.}~\bibnamefont
  {Yamada}}\ and\ \bibinfo {author} {\bibfnamefont {G.}~\bibnamefont
  {Shirane}},\ }\bibfield  {title} {\enquote {\bibinfo {title} {{Neutron
  scattering and nature of the soft optical phonon in SrTiO$_3$}},}\ }\href
  {\doibase 10.1143/JPSJ.26.396} {\bibfield  {journal} {\bibinfo  {journal} {J.
  Phys. Soc. Jpn.}\ }\textbf {\bibinfo {volume} {26}},\ \bibinfo {pages} {396}
  (\bibinfo {year} {1969})}\BibitemShut {NoStop}%
\bibitem [{\citenamefont {Uwe}\ and\ \citenamefont
  {Sakudo}(1976)}]{Hiromoto1976a}%
  \BibitemOpen
  \bibfield  {author} {\bibinfo {author} {\bibfnamefont {H.}~\bibnamefont
  {Uwe}}\ and\ \bibinfo {author} {\bibfnamefont {T.}~\bibnamefont {Sakudo}},\
  }\bibfield  {title} {\enquote {\bibinfo {title} {Stress-induced
  ferroelectricity and soft phonon modes in {SrTiO}$_{3}$},}\ }\href
  {https://link.aps.org/doi/10.1103/PhysRevB.13.271} {\bibfield  {journal}
  {\bibinfo  {journal} {Phys. Rev. B}\ }\textbf {\bibinfo {volume} {13}},\
  \bibinfo {pages} {271--286} (\bibinfo {year} {1976})}\BibitemShut {NoStop}%
\bibitem [{\citenamefont {Itoh}\ \emph {et~al.}(1999)\citenamefont {Itoh},
  \citenamefont {Wang}, \citenamefont {Inaguma}, \citenamefont {Yamaguchi},
  \citenamefont {Shan},\ and\ \citenamefont {Nakamura}}]{Itoh1999a}%
  \BibitemOpen
  \bibfield  {author} {\bibinfo {author} {\bibfnamefont {M.}~\bibnamefont
  {Itoh}}, \bibinfo {author} {\bibfnamefont {R.}~\bibnamefont {Wang}}, \bibinfo
  {author} {\bibfnamefont {Y.}~\bibnamefont {Inaguma}}, \bibinfo {author}
  {\bibfnamefont {T.}~\bibnamefont {Yamaguchi}}, \bibinfo {author}
  {\bibfnamefont {Y-J.}\ \bibnamefont {Shan}}, \ and\ \bibinfo {author}
  {\bibfnamefont {T.}~\bibnamefont {Nakamura}},\ }\bibfield  {title} {\enquote
  {\bibinfo {title} {{Ferroelectricity induced by oxygen isotope exchange in
  strontium titanate perovskite}},}\ }\href
  {https://link.aps.org/doi/10.1103/PhysRevLett.82.3540} {\bibfield  {journal}
  {\bibinfo  {journal} {Phys. Rev. Lett.}\ }\textbf {\bibinfo {volume} {82}},\
  \bibinfo {pages} {3540} (\bibinfo {year} {1999})}\BibitemShut {NoStop}%
\bibitem [{\citenamefont {Bednorz}\ and\ \citenamefont
  {M\"uller}(1984)}]{Bednorz1984a}%
  \BibitemOpen
  \bibfield  {author} {\bibinfo {author} {\bibfnamefont {J.~G.}\ \bibnamefont
  {Bednorz}}\ and\ \bibinfo {author} {\bibfnamefont {K.~A.}\ \bibnamefont
  {M\"uller}},\ }\bibfield  {title} {\enquote {\bibinfo {title}
  {{Sr$_{1-x}$Ca$_x$TiO$_3$: An xy quantum ferroelectric with transition to
  randomness}},}\ }\href {\doibase 10.1103/PhysRevLett.52.2289} {\bibfield
  {journal} {\bibinfo  {journal} {Phys. Rev. Lett.}\ }\textbf {\bibinfo
  {volume} {52}},\ \bibinfo {pages} {2289} (\bibinfo {year}
  {1984})}\BibitemShut {NoStop}%
\bibitem [{\citenamefont {Chandra}\ \emph {et~al.}(2017)\citenamefont
  {Chandra}, \citenamefont {Lonzarich}, \citenamefont {Rowley},\ and\
  \citenamefont {Scott}}]{Chandra2017a}%
  \BibitemOpen
  \bibfield  {author} {\bibinfo {author} {\bibfnamefont {P.}~\bibnamefont
  {Chandra}}, \bibinfo {author} {\bibfnamefont {G.~G.}\ \bibnamefont
  {Lonzarich}}, \bibinfo {author} {\bibfnamefont {S.~E.}\ \bibnamefont
  {Rowley}}, \ and\ \bibinfo {author} {\bibfnamefont {J.~F.}\ \bibnamefont
  {Scott}},\ }\bibfield  {title} {\enquote {\bibinfo {title} {{Prospects and
  applications near ferroelectric quantum phase transitions}},}\ }\href
  {https://doi.org/10.1088/1361-6633/aa82d2} {\bibfield  {journal} {\bibinfo
  {journal} {Rep. Prog. Phys.}\ }\textbf {\bibinfo {volume} {80}},\ \bibinfo
  {pages} {112502} (\bibinfo {year} {2017})}\BibitemShut {NoStop}%
\bibitem [{\citenamefont {Kedem}\ \emph {et~al.}(2016)\citenamefont {Kedem},
  \citenamefont {Zhu},\ and\ \citenamefont {Balatsky}}]{Kedem2016a}%
  \BibitemOpen
  \bibfield  {author} {\bibinfo {author} {\bibfnamefont {Y.}~\bibnamefont
  {Kedem}}, \bibinfo {author} {\bibfnamefont {J.-X.}\ \bibnamefont {Zhu}}, \
  and\ \bibinfo {author} {\bibfnamefont {A.~V.}\ \bibnamefont {Balatsky}},\
  }\bibfield  {title} {\enquote {\bibinfo {title} {Unusual superconducting
  isotope effect in the presence of a quantum criticality},}\ }\href
  {http://link.aps.org/doi/10.1103/PhysRevB.93.184507} {\bibfield  {journal}
  {\bibinfo  {journal} {Phys. Rev. B}\ }\textbf {\bibinfo {volume} {93}},\
  \bibinfo {pages} {184507} (\bibinfo {year} {2016})}\BibitemShut {NoStop}%
\bibitem [{\citenamefont {Dunnett}\ \emph
  {et~al.}(2018{\natexlab{a}})\citenamefont {Dunnett}, \citenamefont {Narayan},
  \citenamefont {Spaldin},\ and\ \citenamefont {Balatsky}}]{Dunnett2016a}%
  \BibitemOpen
  \bibfield  {author} {\bibinfo {author} {\bibfnamefont {K.}~\bibnamefont
  {Dunnett}}, \bibinfo {author} {\bibfnamefont {A.}~\bibnamefont {Narayan}},
  \bibinfo {author} {\bibfnamefont {N.~A.}\ \bibnamefont {Spaldin}}, \ and\
  \bibinfo {author} {\bibfnamefont {A.~V.}\ \bibnamefont {Balatsky}},\
  }\bibfield  {title} {\enquote {\bibinfo {title} {{Strain and ferroelectric
  soft-mode induced superconductivity in strontium titanate}},}\ }\href
  {\doibase 10.1103/PhysRevB.97.144506} {\bibfield  {journal} {\bibinfo
  {journal} {Phys. Rev. B}\ }\textbf {\bibinfo {volume} {97}},\ \bibinfo
  {pages} {144506} (\bibinfo {year} {2018}{\natexlab{a}})}\BibitemShut
  {NoStop}%
\bibitem [{\citenamefont {Lines}\ and\ \citenamefont
  {Glass}(2001)}]{Lines2001a}%
  \BibitemOpen
  \bibfield  {author} {\bibinfo {author} {\bibfnamefont {M.~E.}\ \bibnamefont
  {Lines}}\ and\ \bibinfo {author} {\bibfnamefont {A.~M.}\ \bibnamefont
  {Glass}},\ }\href@noop {} {\emph {\bibinfo {title} {Principles and
  Applications of Ferroelectrics and Related Materials}}}\ (\bibinfo
  {publisher} {Oxford University Press},\ \bibinfo {year} {2001})\BibitemShut
  {NoStop}%
\bibitem [{\citenamefont {Inoue}(1983)}]{Inoue1983a}%
  \BibitemOpen
  \bibfield  {author} {\bibinfo {author} {\bibfnamefont {K.}~\bibnamefont
  {Inoue}},\ }\bibfield  {title} {\enquote {\bibinfo {title} {{Study of
  structural phase transitions by the hyper-Raman scattering}},}\ }\href
  {\doibase 10.1080/00150198308208259} {\bibfield  {journal} {\bibinfo
  {journal} {Ferroelectrics}\ }\textbf {\bibinfo {volume} {52}},\ \bibinfo
  {pages} {253--262} (\bibinfo {year} {1983})}\BibitemShut {NoStop}%
\bibitem [{\citenamefont {Blinc}\ \emph {et~al.}(2005)\citenamefont {Blinc},
  \citenamefont {Zalar}, \citenamefont {Laguta},\ and\ \citenamefont
  {Itoh}}]{Blinc2005a}%
  \BibitemOpen
  \bibfield  {author} {\bibinfo {author} {\bibfnamefont {R.}~\bibnamefont
  {Blinc}}, \bibinfo {author} {\bibfnamefont {B.}~\bibnamefont {Zalar}},
  \bibinfo {author} {\bibfnamefont {V.~V.}\ \bibnamefont {Laguta}}, \ and\
  \bibinfo {author} {\bibfnamefont {M.}~\bibnamefont {Itoh}},\ }\bibfield
  {title} {\enquote {\bibinfo {title} {{Order-disorder component in the phase
  transition mechanism of $^{18}$O Enriched strontium titanate}},}\ }\href
  {https://link.aps.org/doi/10.1103/PhysRevLett.94.147601} {\bibfield
  {journal} {\bibinfo  {journal} {Phys. Rev. Lett.}\ }\textbf {\bibinfo
  {volume} {94}},\ \bibinfo {pages} {147601} (\bibinfo {year}
  {2005})}\BibitemShut {NoStop}%
\bibitem [{\citenamefont {Takesada}\ \emph {et~al.}(2006)\citenamefont
  {Takesada}, \citenamefont {Itoh},\ and\ \citenamefont
  {Yagi}}]{Takesada2006a}%
  \BibitemOpen
  \bibfield  {author} {\bibinfo {author} {\bibfnamefont {M.}~\bibnamefont
  {Takesada}}, \bibinfo {author} {\bibfnamefont {M.}~\bibnamefont {Itoh}}, \
  and\ \bibinfo {author} {\bibfnamefont {T.}~\bibnamefont {Yagi}},\ }\bibfield
  {title} {\enquote {\bibinfo {title} {{Perfect softening of the ferroelectric
  mode in the isotope-exchanged strontium titanate of
  $\mathrm{SrTi}^{18}\mathrm{O}_{3}$ studied by light scattering}},}\ }\href
  {https://link.aps.org/doi/10.1103/PhysRevLett.96.227602} {\bibfield
  {journal} {\bibinfo  {journal} {Phys. Rev. Lett.}\ }\textbf {\bibinfo
  {volume} {96}},\ \bibinfo {pages} {227602} (\bibinfo {year}
  {2006})}\BibitemShut {NoStop}%
\bibitem [{\citenamefont {Taniguchi}\ \emph {et~al.}(2007)\citenamefont
  {Taniguchi}, \citenamefont {Itoh},\ and\ \citenamefont
  {Yagi}}]{Taniguchi2007a}%
  \BibitemOpen
  \bibfield  {author} {\bibinfo {author} {\bibfnamefont {H.}~\bibnamefont
  {Taniguchi}}, \bibinfo {author} {\bibfnamefont {M.}~\bibnamefont {Itoh}}, \
  and\ \bibinfo {author} {\bibfnamefont {T.}~\bibnamefont {Yagi}},\ }\bibfield
  {title} {\enquote {\bibinfo {title} {{Ideal soft mode-type quantum phase
  transition and phase coexistence at quantum critical point in
  $^{18}\mathrm{O}$-Exchanged ${\mathrm{SrTiO}}_{3}$}},}\ }\href {\doibase
  10.1103/PhysRevLett.99.017602} {\bibfield  {journal} {\bibinfo  {journal}
  {Phys. Rev. Lett.}\ }\textbf {\bibinfo {volume} {99}},\ \bibinfo {pages}
  {017602} (\bibinfo {year} {2007})}\BibitemShut {NoStop}%
\bibitem [{\citenamefont {Yagi}\ \emph {et~al.}(2009)\citenamefont {Yagi},
  \citenamefont {Takesada}, \citenamefont {Taniguchi},\ and\ \citenamefont
  {Itoh}}]{Yagi2009a}%
  \BibitemOpen
  \bibfield  {author} {\bibinfo {author} {\bibfnamefont {T.}~\bibnamefont
  {Yagi}}, \bibinfo {author} {\bibfnamefont {M.}~\bibnamefont {Takesada}},
  \bibinfo {author} {\bibfnamefont {M}~\bibnamefont {Taniguchi}}, \ and\
  \bibinfo {author} {\bibfnamefont {M.}~\bibnamefont {Itoh}},\ }\bibfield
  {title} {\enquote {\bibinfo {title} {{Soft-mode dynamics in the ferroelectric
  phase transition of quantum paraelectric SrTiO$_3$}},}\ }\href
  {https://doi.org/10.1080/00150190902852117} {\bibfield  {journal} {\bibinfo
  {journal} {Ferroelectrics}\ }\textbf {\bibinfo {volume} {379}},\ \bibinfo
  {pages} {168} (\bibinfo {year} {2009})}\BibitemShut {NoStop}%
\bibitem [{\citenamefont {Rabe}\ \emph {et~al.}(2007)\citenamefont {Rabe},
  \citenamefont {Ahn},\ and\ \citenamefont {Triscone}}]{ModernPerspective}%
  \BibitemOpen
  \bibinfo {editor} {\bibfnamefont {K.}~\bibnamefont {Rabe}}, \bibinfo {editor}
  {\bibfnamefont {Ch.~H.}\ \bibnamefont {Ahn}}, \ and\ \bibinfo {editor}
  {\bibfnamefont {J.-M.}\ \bibnamefont {Triscone}},\ eds.,\ \href@noop {}
  {\emph {\bibinfo {title} {Physics of Ferroelectrics: A Modern Perspective}}}\
  (\bibinfo  {publisher} {Springer-Verlag},\ \bibinfo {address} {Berlin},\
  \bibinfo {year} {2007})\BibitemShut {NoStop}%
\bibitem [{\citenamefont {Gabay}\ and\ \citenamefont
  {Triscone}(2017)}]{Gabay2017a}%
  \BibitemOpen
  \bibfield  {author} {\bibinfo {author} {\bibfnamefont {M.}~\bibnamefont
  {Gabay}}\ and\ \bibinfo {author} {\bibfnamefont {J.-M.}\ \bibnamefont
  {Triscone}},\ }\bibfield  {title} {\enquote {\bibinfo {title}
  {{Ferroelectricity woos pairing}},}\ }\href
  {http://dx.doi.org/10.1038/nphys4124} {\bibfield  {journal} {\bibinfo
  {journal} {Nat. Phys.}\ }\textbf {\bibinfo {volume} {13}},\ \bibinfo {pages}
  {624} (\bibinfo {year} {2017})}\BibitemShut {NoStop}%
\bibitem [{\citenamefont {Wang}\ \emph {et~al.}(2016)\citenamefont {Wang},
  \citenamefont {{McKeown Walker}}, \citenamefont {Tamai}, \citenamefont
  {Wang}, \citenamefont {Ristic}, \citenamefont {Bruno}, \citenamefont {de~la
  Torre}, \citenamefont {Ricc{\`{o}}}, \citenamefont {Plumb}, \citenamefont
  {Shi}, \citenamefont {Hlawenka}, \citenamefont {S{\'{a}}nchez-Barriga},
  \citenamefont {Varykhalov}, \citenamefont {Kim}, \citenamefont {Hoesch},
  \citenamefont {King}, \citenamefont {Meevasana}, \citenamefont {Diebold},
  \citenamefont {Mesot}, \citenamefont {Moritz}, \citenamefont {Devereaux},
  \citenamefont {Radovic},\ and\ \citenamefont {Baumberger}}]{Wang2016a}%
  \BibitemOpen
  \bibfield  {author} {\bibinfo {author} {\bibfnamefont {Z.}~\bibnamefont
  {Wang}}, \bibinfo {author} {\bibfnamefont {S.}~\bibnamefont {{McKeown
  Walker}}}, \bibinfo {author} {\bibfnamefont {A.}~\bibnamefont {Tamai}},
  \bibinfo {author} {\bibfnamefont {Y.}~\bibnamefont {Wang}}, \bibinfo {author}
  {\bibfnamefont {Z.}~\bibnamefont {Ristic}}, \bibinfo {author} {\bibfnamefont
  {F.~Y.}\ \bibnamefont {Bruno}}, \bibinfo {author} {\bibfnamefont
  {A.}~\bibnamefont {de~la Torre}}, \bibinfo {author} {\bibfnamefont
  {S.}~\bibnamefont {Ricc{\`{o}}}}, \bibinfo {author} {\bibfnamefont {N.~C.}\
  \bibnamefont {Plumb}}, \bibinfo {author} {\bibfnamefont {M.}~\bibnamefont
  {Shi}}, \bibinfo {author} {\bibfnamefont {P.}~\bibnamefont {Hlawenka}},
  \bibinfo {author} {\bibfnamefont {J.}~\bibnamefont {S{\'{a}}nchez-Barriga}},
  \bibinfo {author} {\bibfnamefont {A.}~\bibnamefont {Varykhalov}}, \bibinfo
  {author} {\bibfnamefont {T.~K.}\ \bibnamefont {Kim}}, \bibinfo {author}
  {\bibfnamefont {M.}~\bibnamefont {Hoesch}}, \bibinfo {author} {\bibfnamefont
  {P.~D.~C.}\ \bibnamefont {King}}, \bibinfo {author} {\bibfnamefont
  {W.}~\bibnamefont {Meevasana}}, \bibinfo {author} {\bibfnamefont
  {U.}~\bibnamefont {Diebold}}, \bibinfo {author} {\bibfnamefont
  {J.}~\bibnamefont {Mesot}}, \bibinfo {author} {\bibfnamefont
  {B.}~\bibnamefont {Moritz}}, \bibinfo {author} {\bibfnamefont {T.~P.}\
  \bibnamefont {Devereaux}}, \bibinfo {author} {\bibfnamefont {M.}~\bibnamefont
  {Radovic}}, \ and\ \bibinfo {author} {\bibfnamefont {F.}~\bibnamefont
  {Baumberger}},\ }\bibfield  {title} {\enquote {\bibinfo {title} {{Tailoring
  the nature and strength of electron-phonon interactions in the SrTiO$_3$
  (001) 2D electron liquid}},}\ }\href {\doibase 10.1038/nmat4623} {\bibfield
  {journal} {\bibinfo  {journal} {Nat. Mat.}\ }\textbf {\bibinfo {volume}
  {15}},\ \bibinfo {pages} {835} (\bibinfo {year} {2016})}\BibitemShut
  {NoStop}%
\bibitem [{\citenamefont {van~der Marel}\ \emph {et~al.}(2011)\citenamefont
  {van~der Marel}, \citenamefont {van Mechelen},\ and\ \citenamefont
  {Mazin}}]{Marel2011a}%
  \BibitemOpen
  \bibfield  {author} {\bibinfo {author} {\bibfnamefont {D.}~\bibnamefont
  {van~der Marel}}, \bibinfo {author} {\bibfnamefont {J.~L.~M.}\ \bibnamefont
  {van Mechelen}}, \ and\ \bibinfo {author} {\bibfnamefont {I.~I.}\
  \bibnamefont {Mazin}},\ }\bibfield  {title} {\enquote {\bibinfo {title}
  {{Common Fermi-liquid origin of ${T}^{2}$ resistivity and superconductivity
  in $n$-type SrTiO${}_{3}$}},}\ }\href {\doibase 10.1103/PhysRevB.84.205111}
  {\bibfield  {journal} {\bibinfo  {journal} {Phys. Rev. B}\ }\textbf {\bibinfo
  {volume} {84}},\ \bibinfo {pages} {205111} (\bibinfo {year}
  {2011})}\BibitemShut {NoStop}%
\bibitem [{\citenamefont {Devreese}\ \emph {et~al.}(2010)\citenamefont
  {Devreese}, \citenamefont {Klimin}, \citenamefont {van Mechelen},\ and\
  \citenamefont {van~der Marel}}]{Devreese2010a}%
  \BibitemOpen
  \bibfield  {author} {\bibinfo {author} {\bibfnamefont {J.~T.}\ \bibnamefont
  {Devreese}}, \bibinfo {author} {\bibfnamefont {S.~N.}\ \bibnamefont
  {Klimin}}, \bibinfo {author} {\bibfnamefont {J.~L.~M.}\ \bibnamefont {van
  Mechelen}}, \ and\ \bibinfo {author} {\bibfnamefont {D.}~\bibnamefont
  {van~der Marel}},\ }\bibfield  {title} {\enquote {\bibinfo {title}
  {{Many-body large polaron optical conductivity in
  ${\text{SrTi}}_{1\ensuremath{-}x}{\text{Nb}}_{x}{\text{O}}_{3}$}},}\ }\href
  {\doibase 10.1103/PhysRevB.81.125119} {\bibfield  {journal} {\bibinfo
  {journal} {Phys. Rev. B}\ }\textbf {\bibinfo {volume} {81}},\ \bibinfo
  {pages} {125119} (\bibinfo {year} {2010})}\BibitemShut {NoStop}%
\bibitem [{\citenamefont {Meevasana}\ \emph {et~al.}(2010)\citenamefont
  {Meevasana}, \citenamefont {Zhou}, \citenamefont {Moritz}, \citenamefont
  {Chen}, \citenamefont {He}, \citenamefont {Fujimori}, \citenamefont {Lu},
  \citenamefont {Mo}, \citenamefont {Moore}, \citenamefont {Baumberger},
  \citenamefont {Devereaux}, \citenamefont {van~der Marel}, \citenamefont
  {Nagaosa}, \citenamefont {Zaanen},\ and\ \citenamefont
  {Shen}}]{Meevasana2010a}%
  \BibitemOpen
  \bibfield  {author} {\bibinfo {author} {\bibfnamefont {W.}~\bibnamefont
  {Meevasana}}, \bibinfo {author} {\bibfnamefont {X.~J.}\ \bibnamefont {Zhou}},
  \bibinfo {author} {\bibfnamefont {B.}~\bibnamefont {Moritz}}, \bibinfo
  {author} {\bibfnamefont {C-C.}\ \bibnamefont {Chen}}, \bibinfo {author}
  {\bibfnamefont {R.~H.}\ \bibnamefont {He}}, \bibinfo {author} {\bibfnamefont
  {S-I.}\ \bibnamefont {Fujimori}}, \bibinfo {author} {\bibfnamefont {D.~H.}\
  \bibnamefont {Lu}}, \bibinfo {author} {\bibfnamefont {S-K.}\ \bibnamefont
  {Mo}}, \bibinfo {author} {\bibfnamefont {R.~G.}\ \bibnamefont {Moore}},
  \bibinfo {author} {\bibfnamefont {F.}~\bibnamefont {Baumberger}}, \bibinfo
  {author} {\bibfnamefont {T.~P.}\ \bibnamefont {Devereaux}}, \bibinfo {author}
  {\bibfnamefont {D.}~\bibnamefont {van~der Marel}}, \bibinfo {author}
  {\bibfnamefont {N.}~\bibnamefont {Nagaosa}}, \bibinfo {author} {\bibfnamefont
  {J.}~\bibnamefont {Zaanen}}, \ and\ \bibinfo {author} {\bibfnamefont {Z-X.}\
  \bibnamefont {Shen}},\ }\bibfield  {title} {\enquote {\bibinfo {title}
  {{Strong energy-momentum dispersion of phonon-dressed carriers in the lightly
  doped band insulator SrTiO$_3$}},}\ }\href
  {http://stacks.iop.org/1367-2630/12/i=2/a=023004} {\bibfield  {journal}
  {\bibinfo  {journal} {New J. Phys.}\ }\textbf {\bibinfo {volume} {12}},\
  \bibinfo {pages} {023004} (\bibinfo {year} {2010})}\BibitemShut {NoStop}%
\bibitem [{\citenamefont {van Mechelen}\ \emph {et~al.}(2008)\citenamefont {van
  Mechelen}, \citenamefont {van~der Marel}, \citenamefont {Grimaldi},
  \citenamefont {Kuzmenko}, \citenamefont {Armitage}, \citenamefont {Reyren},
  \citenamefont {Hagemann},\ and\ \citenamefont {Mazin}}]{Mechelen2008a}%
  \BibitemOpen
  \bibfield  {author} {\bibinfo {author} {\bibfnamefont {J.~L.~M.}\
  \bibnamefont {van Mechelen}}, \bibinfo {author} {\bibfnamefont
  {D.}~\bibnamefont {van~der Marel}}, \bibinfo {author} {\bibfnamefont
  {C.}~\bibnamefont {Grimaldi}}, \bibinfo {author} {\bibfnamefont {A.~B.}\
  \bibnamefont {Kuzmenko}}, \bibinfo {author} {\bibfnamefont {N.~P.}\
  \bibnamefont {Armitage}}, \bibinfo {author} {\bibfnamefont {N.}~\bibnamefont
  {Reyren}}, \bibinfo {author} {\bibfnamefont {H.}~\bibnamefont {Hagemann}}, \
  and\ \bibinfo {author} {\bibfnamefont {I.~I.}\ \bibnamefont {Mazin}},\
  }\bibfield  {title} {\enquote {\bibinfo {title} {{Electron-phonon interaction
  and charge carrier mass enhancement in ${\mathrm{SrTiO}}_{3}$}},}\ }\href
  {\doibase 10.1103/PhysRevLett.100.226403} {\bibfield  {journal} {\bibinfo
  {journal} {Phys. Rev. Lett.}\ }\textbf {\bibinfo {volume} {100}},\ \bibinfo
  {pages} {226403} (\bibinfo {year} {2008})}\BibitemShut {NoStop}%
\bibitem [{\citenamefont {Rowley}\ \emph {et~al.}(2018)\citenamefont {Rowley},
  \citenamefont {Enderlein}, \citenamefont {Ferreira~de Oliveira},
  \citenamefont {Tompsett}, \citenamefont {Baggio~Saitovitch}, \citenamefont
  {Saxena},\ and\ \citenamefont {Lonzarich}}]{Rowley2018a}%
  \BibitemOpen
  \bibfield  {author} {\bibinfo {author} {\bibfnamefont {S.~E.}\ \bibnamefont
  {Rowley}}, \bibinfo {author} {\bibfnamefont {C.}~\bibnamefont {Enderlein}},
  \bibinfo {author} {\bibfnamefont {J.}~\bibnamefont {Ferreira~de Oliveira}},
  \bibinfo {author} {\bibfnamefont {D.~A.}\ \bibnamefont {Tompsett}}, \bibinfo
  {author} {\bibfnamefont {E.}~\bibnamefont {Baggio~Saitovitch}}, \bibinfo
  {author} {\bibfnamefont {S.~S.}\ \bibnamefont {Saxena}}, \ and\ \bibinfo
  {author} {\bibfnamefont {G.~G.}\ \bibnamefont {Lonzarich}},\ }\bibfield
  {title} {\enquote {\bibinfo {title} {{Superconductivity in the vicinity of a
  ferroelectric quantum phase transition}},}\ }\href
  {https://arxiv.org/abs/1801.08121} {\bibfield  {journal} {\bibinfo  {journal}
  {arXiv:1801.08121}\ } (\bibinfo {year} {2018})}\BibitemShut {NoStop}%
\bibitem [{\citenamefont {Pfeiffer}\ and\ \citenamefont
  {Schooley}(1970)}]{Pfeiffer1970a}%
  \BibitemOpen
  \bibfield  {author} {\bibinfo {author} {\bibfnamefont {E.~R.}\ \bibnamefont
  {Pfeiffer}}\ and\ \bibinfo {author} {\bibfnamefont {J.~F.}\ \bibnamefont
  {Schooley}},\ }\bibfield  {title} {\enquote {\bibinfo {title} {{Effect of
  stress on the superconducting transition temperature of SrTiO$_3$}},}\ }\href
  {\doibase 10.1007/BF00652506} {\bibfield  {journal} {\bibinfo  {journal} {J.
  Low Temp. Phys.}\ }\textbf {\bibinfo {volume} {2}},\ \bibinfo {pages} {333}
  (\bibinfo {year} {1970})}\BibitemShut {NoStop}%
\bibitem [{\citenamefont {McMillan}(1968)}]{McMillian1968a}%
  \BibitemOpen
  \bibfield  {author} {\bibinfo {author} {\bibfnamefont {W.~L.}\ \bibnamefont
  {McMillan}},\ }\bibfield  {title} {\enquote {\bibinfo {title} {Transition
  temperature of strong-coupled superconductors},}\ }\href
  {https://link.aps.org/doi/10.1103/PhysRev.167.331} {\bibfield  {journal}
  {\bibinfo  {journal} {Phys. Rev.}\ }\textbf {\bibinfo {volume} {167}},\
  \bibinfo {pages} {331} (\bibinfo {year} {1968})}\BibitemShut {NoStop}%
\bibitem [{\citenamefont {Giustino}(2017)}]{Giustino2017a}%
  \BibitemOpen
  \bibfield  {author} {\bibinfo {author} {\bibfnamefont {F.}~\bibnamefont
  {Giustino}},\ }\bibfield  {title} {\enquote {\bibinfo {title}
  {Electron-phonon interactions from first principles},}\ }\href
  {https://link.aps.org/doi/10.1103/RevModPhys.89.015003} {\bibfield  {journal}
  {\bibinfo  {journal} {Rev. Mod. Phys.}\ }\textbf {\bibinfo {volume} {89}},\
  \bibinfo {pages} {015003} (\bibinfo {year} {2017})}\BibitemShut {NoStop}%
\bibitem [{\citenamefont {W\"olfle}\ and\ \citenamefont
  {Balatsky}(2018)}]{Woelfe2018a}%
  \BibitemOpen
  \bibfield  {author} {\bibinfo {author} {\bibfnamefont {P.}~\bibnamefont
  {W\"olfle}}\ and\ \bibinfo {author} {\bibfnamefont {A.~V.}\ \bibnamefont
  {Balatsky}},\ }\bibfield  {title} {\enquote {\bibinfo {title}
  {{Superconductivity at low density near a ferroelectric quantum critical
  point: Doped SrTiO$_{3}$}},}\ }\href
  {https://link.aps.org/doi/10.1103/PhysRevB.98.104505} {\bibfield  {journal}
  {\bibinfo  {journal} {Phys. Rev. B}\ }\textbf {\bibinfo {volume} {98}},\
  \bibinfo {pages} {104505} (\bibinfo {year} {2018})}\BibitemShut {NoStop}%
\bibitem [{\citenamefont {Pytte}(1972)}]{Pytte1972a}%
  \BibitemOpen
  \bibfield  {author} {\bibinfo {author} {\bibfnamefont {E.}~\bibnamefont
  {Pytte}},\ }\bibfield  {title} {\enquote {\bibinfo {title} {{Theory of
  perovskite ferroelectrics}},}\ }\href {\doibase 10.1103/PhysRevB.5.3758}
  {\bibfield  {journal} {\bibinfo  {journal} {Phys. Rev. B}\ }\textbf {\bibinfo
  {volume} {5}},\ \bibinfo {pages} {3758} (\bibinfo {year} {1972})}\BibitemShut
  {NoStop}%
\bibitem [{\citenamefont {Aharony}\ and\ \citenamefont
  {Fisher}(1973)}]{Aharony1973a}%
  \BibitemOpen
  \bibfield  {author} {\bibinfo {author} {\bibfnamefont {A.}~\bibnamefont
  {Aharony}}\ and\ \bibinfo {author} {\bibfnamefont {M.~E.}\ \bibnamefont
  {Fisher}},\ }\bibfield  {title} {\enquote {\bibinfo {title} {{Critical
  behavior of magnets with dipolar Interactions. I. Renormalization group near
  four dimensions}},}\ }\href {\doibase 10.1103/PhysRevB.8.3323} {\bibfield
  {journal} {\bibinfo  {journal} {Phys. Rev. B}\ }\textbf {\bibinfo {volume}
  {8}},\ \bibinfo {pages} {3323} (\bibinfo {year} {1973})}\BibitemShut
  {NoStop}%
\bibitem [{\citenamefont {Arce-Gamboa}\ and\ \citenamefont
  {Guzm{\'{a}}n-Verri}(2017)}]{Arce-Gamboa2017a}%
  \BibitemOpen
  \bibfield  {author} {\bibinfo {author} {\bibfnamefont {J.~R.}\ \bibnamefont
  {Arce-Gamboa}}\ and\ \bibinfo {author} {\bibfnamefont {G.~G.}\ \bibnamefont
  {Guzm{\'{a}}n-Verri}},\ }\bibfield  {title} {\enquote {\bibinfo {title}
  {{Random electric field instabilities of relaxor ferroelectrics}},}\ }\href
  {\doibase 10.1038/s41535-017-0032-9} {\bibfield  {journal} {\bibinfo
  {journal} {npj Quantum Materials}\ }\textbf {\bibinfo {volume} {2}},\
  \bibinfo {pages} {28} (\bibinfo {year} {2017})}\BibitemShut {NoStop}%
\bibitem [{\citenamefont {Strukov}\ and\ \citenamefont
  {Levanyuk}(1998)}]{LevanyukFerroelectricity}%
  \BibitemOpen
  \bibfield  {author} {\bibinfo {author} {\bibfnamefont {B.~A.}\ \bibnamefont
  {Strukov}}\ and\ \bibinfo {author} {\bibfnamefont {A.~P.}\ \bibnamefont
  {Levanyuk}},\ }\href@noop {} {\emph {\bibinfo {title} {Ferroelectric
  Phenomena in Crystals: Physical Foundations}}}\ (\bibinfo  {publisher}
  {Springer},\ \bibinfo {address} {Berlin},\ \bibinfo {year}
  {1998})\BibitemShut {NoStop}%
\bibitem [{\citenamefont {Zhong}\ and\ \citenamefont
  {Vanderbilt}(1996)}]{Zhong1996a}%
  \BibitemOpen
  \bibfield  {author} {\bibinfo {author} {\bibfnamefont {W.}~\bibnamefont
  {Zhong}}\ and\ \bibinfo {author} {\bibfnamefont {D.}~\bibnamefont
  {Vanderbilt}},\ }\bibfield  {title} {\enquote {\bibinfo {title} {{Effect of
  quantum fluctuations on structural phase transitions in
  ${\mathrm{SrTiO}}_{3}$ and ${\mathrm{BaTiO}}_{3}$}},}\ }\href
  {https://link.aps.org/doi/10.1103/PhysRevB.53.5047} {\bibfield  {journal}
  {\bibinfo  {journal} {Phys. Rev. B}\ }\textbf {\bibinfo {volume} {53}},\
  \bibinfo {pages} {5047} (\bibinfo {year} {1996})}\BibitemShut {NoStop}%
\bibitem [{\citenamefont {Cowley}(1980)}]{Cowley1980a}%
  \BibitemOpen
  \bibfield  {author} {\bibinfo {author} {\bibfnamefont {R.~A.}\ \bibnamefont
  {Cowley}},\ }\bibfield  {title} {\enquote {\bibinfo {title} {{Structural
  phase transitions I. Landau theory}},}\ }\href
  {https://doi.org/10.1080/00018738000101346} {\bibfield  {journal} {\bibinfo
  {journal} {Adv. Phys.}\ }\textbf {\bibinfo {volume} {29}},\ \bibinfo {pages}
  {1} (\bibinfo {year} {1980})}\BibitemShut {NoStop}%
\bibitem [{\citenamefont {Samara}(1971)}]{Samara1971a}%
  \BibitemOpen
  \bibfield  {author} {\bibinfo {author} {\bibfnamefont {G.~A.}\ \bibnamefont
  {Samara}},\ }\bibfield  {title} {\enquote {\bibinfo {title} {{The
  Gr\"{u}neisen parameter of the soft ferroelectric mode in the cubic
  perovskites}},}\ }\href {\doibase 10.1080/00150197108241505} {\bibfield
  {journal} {\bibinfo  {journal} {Ferroelectrics}\ }\textbf {\bibinfo {volume}
  {2}},\ \bibinfo {pages} {177} (\bibinfo {year} {1971})}\BibitemShut {NoStop}%
\bibitem [{\citenamefont {Carpenter}(2007)}]{Carpenter2007a}%
  \BibitemOpen
  \bibfield  {author} {\bibinfo {author} {\bibfnamefont {M.}~\bibnamefont
  {Carpenter}},\ }\bibfield  {title} {\enquote {\bibinfo {title} {{Elastic
  anomalies accompanying phase transitions in (Ca,Sr)TiO$_3$ perovskites: Part
  I. Landau theory and a calibration for SrTiO$_3$}},}\ }\href
  {http://dx.doi.org/10.2138/am.2007.2295} {\bibfield  {journal} {\bibinfo
  {journal} {Am. Mineral.}\ }\textbf {\bibinfo {volume} {92}},\ \bibinfo
  {pages} {309} (\bibinfo {year} {2007})}\BibitemShut {NoStop}%
\bibitem [{\citenamefont {Pai}\ \emph {et~al.}(2018)\citenamefont {Pai},
  \citenamefont {Tylan-Tyler}, \citenamefont {Irwin},\ and\ \citenamefont
  {Levy}}]{Pai2018a}%
  \BibitemOpen
  \bibfield  {author} {\bibinfo {author} {\bibfnamefont {Y.-Y.}\ \bibnamefont
  {Pai}}, \bibinfo {author} {\bibfnamefont {A.}~\bibnamefont {Tylan-Tyler}},
  \bibinfo {author} {\bibfnamefont {P.}~\bibnamefont {Irwin}}, \ and\ \bibinfo
  {author} {\bibfnamefont {J.}~\bibnamefont {Levy}},\ }\bibfield  {title}
  {\enquote {\bibinfo {title} {{Physics of SrTiO$_3$-based heterostructures and
  nanostructures: a review}},}\ }\href
  {http://stacks.iop.org/0034-4885/81/i=3/a=036503} {\bibfield  {journal}
  {\bibinfo  {journal} {Rep. Prog. Phys.}\ }\textbf {\bibinfo {volume} {81}},\
  \bibinfo {pages} {036503} (\bibinfo {year} {2018})}\BibitemShut {NoStop}%
\bibitem [{\citenamefont {Gariglio}\ \emph {et~al.}(2016)\citenamefont
  {Gariglio}, \citenamefont {Gabay},\ and\ \citenamefont
  {Triscone}}]{Gariglio2016a}%
  \BibitemOpen
  \bibfield  {author} {\bibinfo {author} {\bibfnamefont {S.}~\bibnamefont
  {Gariglio}}, \bibinfo {author} {\bibfnamefont {M.}~\bibnamefont {Gabay}}, \
  and\ \bibinfo {author} {\bibfnamefont {J.-M.}\ \bibnamefont {Triscone}},\
  }\bibfield  {title} {\enquote {\bibinfo {title} {{Research update:
  conductivity and beyond at the LaAlO$_3$/SrTiO$_3$ interface}},}\ }\href
  {\doibase 10.1063/1.4953822} {\bibfield  {journal} {\bibinfo  {journal} {APL
  Materials}\ }\textbf {\bibinfo {volume} {4}},\ \bibinfo {pages} {060701}
  (\bibinfo {year} {2016})}\BibitemShut {NoStop}%
\bibitem [{\citenamefont {Ueno}\ \emph {et~al.}(2011)\citenamefont {Ueno},
  \citenamefont {Nakamura}, \citenamefont {Shimotani}, \citenamefont {Yuan},
  \citenamefont {Kimura}, \citenamefont {Nojima}, \citenamefont {Aoki},
  \citenamefont {Iwasa},\ and\ \citenamefont {Kawasaki}}]{Ueno2011a}%
  \BibitemOpen
  \bibfield  {author} {\bibinfo {author} {\bibfnamefont {K.}~\bibnamefont
  {Ueno}}, \bibinfo {author} {\bibfnamefont {S.}~\bibnamefont {Nakamura}},
  \bibinfo {author} {\bibfnamefont {H.}~\bibnamefont {Shimotani}}, \bibinfo
  {author} {\bibfnamefont {H.~T.}\ \bibnamefont {Yuan}}, \bibinfo {author}
  {\bibfnamefont {N.}~\bibnamefont {Kimura}}, \bibinfo {author} {\bibfnamefont
  {T.}~\bibnamefont {Nojima}}, \bibinfo {author} {\bibfnamefont
  {H.}~\bibnamefont {Aoki}}, \bibinfo {author} {\bibfnamefont {Y.}~\bibnamefont
  {Iwasa}}, \ and\ \bibinfo {author} {\bibfnamefont {M.}~\bibnamefont
  {Kawasaki}},\ }\bibfield  {title} {\enquote {\bibinfo {title} {{Discovery of
  superconductivity in KTaO$_3$ by electrostatic carrier doping}},}\ }\href
  {\doibase 10.1038/nnano.2011.78} {\bibfield  {journal} {\bibinfo  {journal}
  {Nat. Nanotech.}\ }\textbf {\bibinfo {volume} {6}},\ \bibinfo {pages} {408}
  (\bibinfo {year} {2011})}\BibitemShut {NoStop}%
\bibitem [{\citenamefont {Lee}\ \emph {et~al.}(2014)\citenamefont {Lee},
  \citenamefont {Schmitt}, \citenamefont {Moore}, \citenamefont {Johnston},
  \citenamefont {Cui}, \citenamefont {Li}, \citenamefont {Yi}, \citenamefont
  {Liu}, \citenamefont {Hashimoto}, \citenamefont {Zhang}, \citenamefont {Lu},
  \citenamefont {Devereaux}, \citenamefont {Lee},\ and\ \citenamefont
  {Shen}}]{Lee2014a}%
  \BibitemOpen
  \bibfield  {author} {\bibinfo {author} {\bibfnamefont {J.~J.}\ \bibnamefont
  {Lee}}, \bibinfo {author} {\bibfnamefont {F.~T.}\ \bibnamefont {Schmitt}},
  \bibinfo {author} {\bibfnamefont {R.~G.}\ \bibnamefont {Moore}}, \bibinfo
  {author} {\bibfnamefont {S.}~\bibnamefont {Johnston}}, \bibinfo {author}
  {\bibfnamefont {Y.-T.}\ \bibnamefont {Cui}}, \bibinfo {author} {\bibfnamefont
  {W.}~\bibnamefont {Li}}, \bibinfo {author} {\bibfnamefont {M.}~\bibnamefont
  {Yi}}, \bibinfo {author} {\bibfnamefont {Z.~K.}\ \bibnamefont {Liu}},
  \bibinfo {author} {\bibfnamefont {M.}~\bibnamefont {Hashimoto}}, \bibinfo
  {author} {\bibfnamefont {Y.}~\bibnamefont {Zhang}}, \bibinfo {author}
  {\bibfnamefont {D.~H.}\ \bibnamefont {Lu}}, \bibinfo {author} {\bibfnamefont
  {T.~P.}\ \bibnamefont {Devereaux}}, \bibinfo {author} {\bibfnamefont {D.-H.}\
  \bibnamefont {Lee}}, \ and\ \bibinfo {author} {\bibfnamefont {Z.-X.}\
  \bibnamefont {Shen}},\ }\bibfield  {title} {\enquote {\bibinfo {title}
  {{Interfacial mode coupling as the origin of the enhancement of $T_c$ in FeSe
  films on SrTiO$_3$}},}\ }\href {http://dx.doi.org/10.1038/nature13894}
  {\bibfield  {journal} {\bibinfo  {journal} {Nature}\ }\textbf {\bibinfo
  {volume} {515}},\ \bibinfo {pages} {245} (\bibinfo {year}
  {2014})}\BibitemShut {NoStop}%
\bibitem [{\citenamefont {Morice}\ \emph {et~al.}(2017)\citenamefont {Morice},
  \citenamefont {Chandra}, \citenamefont {Rowley}, \citenamefont {Lonzarich},\
  and\ \citenamefont {Saxena}}]{Morice2017a}%
  \BibitemOpen
  \bibfield  {author} {\bibinfo {author} {\bibfnamefont {C.}~\bibnamefont
  {Morice}}, \bibinfo {author} {\bibfnamefont {P.}~\bibnamefont {Chandra}},
  \bibinfo {author} {\bibfnamefont {S.~E.}\ \bibnamefont {Rowley}}, \bibinfo
  {author} {\bibfnamefont {G.}~\bibnamefont {Lonzarich}}, \ and\ \bibinfo
  {author} {\bibfnamefont {S.~S.}\ \bibnamefont {Saxena}},\ }\bibfield  {title}
  {\enquote {\bibinfo {title} {Hidden fluctuations close to a quantum
  bicritical point},}\ }\href {\doibase 10.1103/PhysRevB.96.245104} {\bibfield
  {journal} {\bibinfo  {journal} {Phys. Rev. B}\ }\textbf {\bibinfo {volume}
  {96}},\ \bibinfo {pages} {245104} (\bibinfo {year} {2017})}\BibitemShut
  {NoStop}%
\bibitem [{\citenamefont {Narayan}\ \emph {et~al.}(2017)\citenamefont
  {Narayan}, \citenamefont {Balatsky},\ and\ \citenamefont
  {Spaldin}}]{Narayan2017a}%
  \BibitemOpen
  \bibfield  {author} {\bibinfo {author} {\bibfnamefont {A.}~\bibnamefont
  {Narayan}}, \bibinfo {author} {\bibfnamefont {A.~V.}\ \bibnamefont
  {Balatsky}}, \ and\ \bibinfo {author} {\bibfnamefont {N.~A.}\ \bibnamefont
  {Spaldin}},\ }\bibfield  {title} {\enquote {\bibinfo {title} {{Multiferroic
  quantum criticality}},}\ }\href {https://arxiv.org/abs/1711.07989} {\bibfield
   {journal} {\bibinfo  {journal} {arXiv:1711.07989}\ } (\bibinfo {year}
  {2017})}\BibitemShut {NoStop}%
\bibitem [{\citenamefont {Dunnett}\ \emph
  {et~al.}(2018{\natexlab{b}})\citenamefont {Dunnett}, \citenamefont {Zhu},
  \citenamefont {Spaldin}, \citenamefont {Juricic},\ and\ \citenamefont
  {Balatsky}}]{Dunnett2018b}%
  \BibitemOpen
  \bibfield  {author} {\bibinfo {author} {\bibfnamefont {K.}~\bibnamefont
  {Dunnett}}, \bibinfo {author} {\bibfnamefont {J.~X.}\ \bibnamefont {Zhu}},
  \bibinfo {author} {\bibfnamefont {N.~A.}\ \bibnamefont {Spaldin}}, \bibinfo
  {author} {\bibfnamefont {V.}~\bibnamefont {Juricic}}, \ and\ \bibinfo
  {author} {\bibfnamefont {A.~V.}\ \bibnamefont {Balatsky}},\ }\bibfield
  {title} {\enquote {\bibinfo {title} {{Dynamic multiferroicity of a
  ferroelectric quantum critical point}},}\ }\href
  {http://arxiv.org/abs/1808.05509} {\bibfield  {journal} {\bibinfo  {journal}
  {arXiv:1808.05509}\ } (\bibinfo {year} {2018}{\natexlab{b}})}\BibitemShut
  {NoStop}%
\bibitem [{\citenamefont {Handunkanda}\ \emph {et~al.}(2015)\citenamefont
  {Handunkanda}, \citenamefont {Curry}, \citenamefont {Voronov}, \citenamefont
  {Said}, \citenamefont {Guzm\'an-Verri}, \citenamefont {Brierley},
  \citenamefont {Littlewood},\ and\ \citenamefont
  {Hancock}}]{Handunkanda2015a}%
  \BibitemOpen
  \bibfield  {author} {\bibinfo {author} {\bibfnamefont {S.~U.}\ \bibnamefont
  {Handunkanda}}, \bibinfo {author} {\bibfnamefont {E.~B.}\ \bibnamefont
  {Curry}}, \bibinfo {author} {\bibfnamefont {V.}~\bibnamefont {Voronov}},
  \bibinfo {author} {\bibfnamefont {A.~H.}\ \bibnamefont {Said}}, \bibinfo
  {author} {\bibfnamefont {G.~G.}\ \bibnamefont {Guzm\'an-Verri}}, \bibinfo
  {author} {\bibfnamefont {R.~T.}\ \bibnamefont {Brierley}}, \bibinfo {author}
  {\bibfnamefont {P.~B.}\ \bibnamefont {Littlewood}}, \ and\ \bibinfo {author}
  {\bibfnamefont {J.~N.}\ \bibnamefont {Hancock}},\ }\bibfield  {title}
  {\enquote {\bibinfo {title} {Large isotropic negative thermal expansion above
  a structural quantum phase transition},}\ }\href
  {https://link.aps.org/doi/10.1103/PhysRevB.92.134101} {\bibfield  {journal}
  {\bibinfo  {journal} {Phys. Rev. B}\ }\textbf {\bibinfo {volume} {92}},\
  \bibinfo {pages} {134101} (\bibinfo {year} {2015})}\BibitemShut {NoStop}%
\bibitem [{\citenamefont {Occhialini}\ \emph {et~al.}(2017)\citenamefont
  {Occhialini}, \citenamefont {Handunkanda}, \citenamefont {Said},
  \citenamefont {Trivedi}, \citenamefont {Guzm\'an-Verri},\ and\ \citenamefont
  {Hancock}}]{Occhialini2017a}%
  \BibitemOpen
  \bibfield  {author} {\bibinfo {author} {\bibfnamefont {C.~A.}\ \bibnamefont
  {Occhialini}}, \bibinfo {author} {\bibfnamefont {S.~U.}\ \bibnamefont
  {Handunkanda}}, \bibinfo {author} {\bibfnamefont {A.}~\bibnamefont {Said}},
  \bibinfo {author} {\bibfnamefont {S.}~\bibnamefont {Trivedi}}, \bibinfo
  {author} {\bibfnamefont {G.~G.}\ \bibnamefont {Guzm\'an-Verri}}, \ and\
  \bibinfo {author} {\bibfnamefont {J.~N.}\ \bibnamefont {Hancock}},\
  }\bibfield  {title} {\enquote {\bibinfo {title} {Negative thermal expansion
  near two structural quantum phase transitions},}\ }\href
  {https://link.aps.org/doi/10.1103/PhysRevMaterials.1.070603} {\bibfield
  {journal} {\bibinfo  {journal} {Phys. Rev. Materials}\ }\textbf {\bibinfo
  {volume} {1}},\ \bibinfo {pages} {070603} (\bibinfo {year}
  {2017})}\BibitemShut {NoStop}%
\end{thebibliography}

\end{document}